\documentclass[%
reprint,
superscriptaddress,
amsmath,amssymb,longbibliography
prb,
]{revtex4-2}

\usepackage{graphicx}
\usepackage{dcolumn}
\usepackage{siunitx}
\usepackage{bm}
\usepackage{hyperref}
\hypersetup{colorlinks=true,linkcolor=blue,citecolor=blue,urlcolor=blue}

\begin{document}

\title{Coexistence of Dirac fermion and charge density wave in square-net-based semimetal \texorpdfstring{LaAuSb$_2$}{}}

\author{Xueliang Wu}
\thanks{These authors contributed equally to this work.}
\affiliation{Low Temperature Physics Laboratory, College of Physics and Center of Quantum Materials and Devices, Chongqing University, Chongqing 401331, China.
}

\author{Zhixiang Hu}
\thanks{These authors contributed equally to this work.}
\affiliation{Condensed Matter Physics and Materials Science Department, Brookhaven National Laboratory, Upton, New York 11973, USA.
}
\affiliation{Department of Material Science and Chemical Engineering, Stony Brook University, Stony Brook, New York 11790, USA.
}

\author{David Graf}
\affiliation{National High Magnetic Field Laboratory, Florida State University, Tallahassee, Florida 32306-4005, USA
}

\author{Yu Liu}
\affiliation{Condensed Matter Physics and Materials Science Department, Brookhaven National Laboratory, Upton, New York 11973, USA.
}
\affiliation{Center for Correlated Matter and School of Physics, Zhejiang University, Hangzhou 310058, P.R. China. 
}

\author{Chaoyue Deng}
\affiliation{College of Physics and Center of Quantum Materials and Devices, Chongqing University, Chongqing 401331, China.}

\author{Huixia Fu}
\affiliation{College of Physics and Center of Quantum Materials and Devices, Chongqing University, Chongqing 401331, China.}

\author{Asish K. Kundu}
\affiliation{Condensed Matter Physics and Materials Science Department, Brookhaven National Laboratory, Upton, New York 11973, USA.
} 

\author{Tonica Valla}
\affiliation{Donostia International Physics Center, Donostia - San Sebastián, 20018, Spain
} 

\author{Cedomir Petrovic}
\email{cedomir.petrovic@stonybrook.edu}
\affiliation{Condensed Matter Physics and Materials Science Department, Brookhaven National Laboratory, Upton, New York 11973, USA.
}
\affiliation{Department of Material Science and Chemical Engineering, Stony Brook University, Stony Brook, New York 11790, USA.
}
\affiliation{Shanghai Advanced Research in Physical Sciences, Shanghai 201203, China.
}
\affiliation{Department of Nuclear and Plasma Physics,
Vinca Institute of Nuclear Sciences, University of Belgrade, Belgrade 11001, Serbia.
}

\author{Aifeng Wang}
\email{afwang@cqu.edu.cn}
\affiliation{Low Temperature Physics Laboratory, College of Physics and Center of Quantum Materials and Devices, Chongqing University, Chongqing 401331, China.
}
\affiliation{Condensed Matter Physics and Materials Science Department, Brookhaven National Laboratory, Upton, New York 11973, USA.
}

\date{\today}

\begin{abstract}
We report a comprehensive study of magnetotransport properties, angle-resolved photoemission spectroscopy (ARPES), and density functional theory (DFT) calculations on self-flux grown LaAuSb$_2$ single crystals. Resistivity and Hall measurements reveal a charge density wave (CDW) transition at 77 K. MR and de Haas-Van Alphen (dHvA) measurements indicate that the transport properties of LaAuSb$_2$ are dominated by Dirac fermions that arise from Sb square nets. ARPES measurements and DFT calculations reveal an electronic structure with a common feature of the square-net-based topological semimetals, which is in good agreement with the magnetotransport properties. Our results indicate the coexistence of CDW and Dirac fermion in LaAuSb$_2$, both of which are linked to the bands arising from the Sb-square net, suggesting that the square net could serve as a structural motif to explore various electronic orders.   
\end{abstract}
\maketitle

\section{INTRODUCTION}
Ternary square-net-based compounds La$T$Sb$_2$ ($T$ = Cu, Ag, Au) have recently attracted renewed interest due to the observation of anisotropic Dirac fermion, charge density wave (CDW), and superconductivity, suggesting that La$T$Sb$_2$ can serve as a new candidate for scrutinizing the intertwined orders \cite{song_charge-density-wave_2003,akiba_observation_2022,shi_observation_2016}. For example, in the typical compound LaAgSb$_2$, two successive CDW transitions were observed at $T_\mathrm{CDW1} =$ 207 K and $T_\mathrm{CDW2} =$ 184 K, respectively \cite{song_charge-density-wave_2003}. ARPES, NMR, and inelastic x-ray scattering (IXS) measurements indicate that CDW in LaAgSb$_2$ is driven by the nesting of Dirac-like bands \cite{shi_observation_2016,baek_nmr_2022,bosak_evidence_2021}. Dirac-like bands are theoretically predicted to originate from the band folding of the Sb-square net and have been experimentally verified by ARPES and magnetotransport measurements \cite{shi_observation_2016,wang_multiband_2012,Akiba_Magnetotransport_2022}. Superconductivity at $T_c =$ 0.3 K was observed at ambient pressure and can be increased to $T_c =$ 1 K by high pressure \cite{akiba_observation_2022}. The theoretical calculation demonstrates that the Sb-square net is also crucial for the emergence of superconductivity \cite{akiba_observation_2022}. These investigations indicate that the Sb square net is the key structural motif responsible for the Dirac dispersion, CDW, and superconductivity \cite{shi_observation_2016,song_charge-density-wave_2003,akiba_observation_2022}.

La$T$Sb$_2$ crystallizes in a tetragonal HfCuSi$_2$-type structure (P4/nmm, No.129) featured by alternatively stacking layers of the Sb-square net, $T$Sb, and La along the $c$ axis. Interestingly, it has been theoretically and experimentally demonstrated that the electronic structure associated with the Sb square net is tunable by its chemical environment, that is, the structural configuration or magnetism of adjacent layers \cite{klemenz_topological_2019,xia_coupling_2023}. In fact, the physical properties, including the lattice parameters, the CDW transition temperature ($T_{\mathrm{CDW}}$), and the superconducting transition temperature ($T_c$) vary systematically in the sequence of Cu, Au, and Ag. For example, LaCuSb$_2$ shows superconductivity below $T_c$ = 0.9 K without CDW transition \cite{muro_magnetic_1997,chamorro_dirac_2019,akiba_phonon-mediated_2023}, LaAuSb$_2$ shows the CDW transition at 88 K and superconductivity at 0.64 K \cite{du_interplay_2020}, LaAgSb$_2$ shows two CDW transitions at 207 K and 184 K \cite{song_charge-density-wave_2003} and superconductivity at 0.3 K \cite{akiba_observation_2022}. Although the relationship between CDW and superconductivity in LaAuSb$_2$ has been investigated by means of doping, magnetotransport, and pressure measurements \cite{kuo_characterization_2019,xiang_tuning_2020, du_interplay_2020,xiang_effects_2022}, the electronic structure and topological properties of LaAuSb$_2$ have been largely unexplored. Therefore, it is of great interest to characterize the electronic structure and physical properties of LaAuSb$_2$, which could deepen our understanding of the intertwined orders associated with the Sb-square net and their interaction with the neighboring layers.

In this paper, we study the electronic structure and physical properties of LaAuSb$_2$ single crystals via density functional theory (DFT) calculations, magnetotransport, and ARPES measurements. The in-plane resistivity reveals a CDW transition at $T_{\mathrm{CDW}} =$ 77 K. A remarkable magnetoresistance (MR) develops below $T_{\mathrm{CDW}}$ and reaches the maximum value of 16\% at 2 K and 9 T. However, no anomaly corresponding to the CDW transition was found in the specific heat data, suggesting that the Fermi surface gap and structural distortion induced by the CDW transition are weak. Kohler's rule is violated in LaAuSb$_2$, which can be attributed to the multiband effects, especially in the CDW state with the reconstructed Fermi surfaces. Hall measurements indicate that the transport properties are dominated by the hole-type carriers, and the carrier concentration is significantly suppressed by the formation of the CDW order. Quantum oscillation was probed by two methods, i.e., magnetization measurements up to 9 T and torque measurements up to 35 T, which are consistent with each other. From the fast Fourier transform (FFT) spectra for the de Haas-Van Alphen (dHvA) oscillation data,  $F$ = 10.9, 54.7, 1021.4, and 1702.3 T can be obtained. Further analysis of dHvA oscillation data reveals that the Fermi pocket corresponding to $F =$ 54.7 T has a quasi-two-dimensional (2D) character with a small effective mass and nontrivial Berry phase, indicating the existence of Dirac fermion. Further, we directly probe the band structure using ARPES measurements. Both the quantum oscillation and ARPES data can be well interpreted by the electronic structure calculated by the density functional theory, where the Dirac bands can be identified along the $\Gamma-M$ line and $X$ point in the Brillouin zone. Hence, our study indicates that the Dirac states are indeed present in the CDW compound LaAuSb$_2$. 

\section{METHODS}
\subsection{Experimental details}
Single crystals of LaAuSb$_2$ were grown by the self-flux method with excess Au and Sb as flux \cite{myers_haasvan_1999}. Mixtures of La chunks, Au wires,  and Sb lumps in the ratio of La$_{0.045}$Au$_{0.091}$Sb$_{0.864}$ were loaded into an alumina crucible and then sealed in an evacuated quartz tube. The sealed quartz tubes were slowly heated to \SI{1180}{\degreeCelsius}, held for 2 h, and then cooled to \SI{670}{\degreeCelsius} at a rate of \SI{3}{\degreeCelsius}/h to grow single crystals. Shiny centimeter-sized single crystals can be obtained by decanting the excess flux using a centrifuge. 

Powder XRD data were collected using a PANalytical powder diffractometer (Cu $K_{\alpha}$ radiation). The element analysis was performed using energy-dispersive x-ray spectroscopy (EDX) in a Thermo Fisher Quattro S Environmental scanning electron microscope (SEM). The instrumental error for La, Au, and Sb is 8\%, 12\%, and 6\%, respectively. Specific heat and magnetotransport measurements up to 9 T were conducted on polished crystals to remove Sb flux droplets on the surface in a Quantum Design DynaCool Physical Properties Measurement System (PPMS-9T). Magnetization was measured using the Quantum Design VSM option in PPMS-9T. The magnetic torque was measured using a piezoresistive cantilever in a 35 T resistive magnet at the National High Magnetic Field Laboratory (NHMFL) in Tallahassee. Electrical resistivity $\rho_{xx}$ was measured using a standard four-probe method. Hall resistivity $\rho_{yx}$ was measured using a four-terminal technique by switching the polarity of the magnetic field to eliminate contributions from $\rho_{xx}$. The ARPES experiments were performed at the Electron Spectro-Microscopy (ESM) 21-ID-1 beamline of the National Synchrotron Light Source II. The beamline is equipped with a Scienta DA30 electron analyzer, with base pressure $\sim2\times10^{-11}$ mbar. The total energy and the angular resolution was $\sim 15$ meV and $\sim$ \ang{0.1}, respectively.

Magnetization, specific heat, and magnetotransport measurements (including in-plane resistivity, Hall resistivity, and magnetoresistance) were performed on a bar-shaped single crystal with a dimension of $2 \times 1 \times 0.2$ mm$^3$ and a weight of 1.1 mg. The out-of-plane resistivity, magnetic torque, and ARPES measurements were performed on different crystals from the same batch. The out-of-plane resistivity sample is prepared by cutting the crystals into a rod with a dimension of $0.5 \times 0.5 \times 0.8$ mm$^3$, where the long side of the rod is along the crystallographic $c$ axis. A small crystal with a dimension of $0.6 \times 0.5 \times 0.2$ mm$^3$ is used for magnetic torque measurements. The dimension of the crystal used for ARPES measurements is about $2 \times 2 \times 0.5$ mm$^3$.We would like to point out that the physical properties inferred from various measurements are consistent with each other, and no signature of inhomogeneity was found.

\subsection{Theoretical methods}
The first-principles calculations were performed using the Vienna $ab$-initio simulation package (VASP) \cite{kresse_efficient_1996,kresse_efficiency_1996} in the framework of density functional theory (DFT). The generalized gradient approximation (GGA) of Perdew-Burke-Ernzerhof (PBE) \cite{perdew_generalized_1996} was implemented for the exchange-correlation potential. The projected-augmented wave (PAW) pseudopotential \cite{blochl_projector_1994} was adopted with an energy cutoff of 400 eV as the basis set. A tetragonal cell in $P4/nmm$ space group with the experimental lattice constants was taken. In structural optimizations, all atoms were fully relaxed within the energy convergence threshold of $10^{-6}$ eV until the residual force on each relaxed atom was less than 0.001 eV/{\AA}. A $\Gamma$-centered Monkhorst-Pack $k$-point mesh of $28 \times 28 \times 12$ was adopted for sampling the first Brillouin zone. To emphasize the effects of spin-orbital coupling (SOC), the electronic properties of LaAuSb$_2$, including band structure,  density of states (DOS), and Fermi surface were obtained with SOC taken into account. The analysis and visualization of the Fermi surfaces were implemented via IFermi software \cite{ganose_ifermi_2021}.

\section{RESULTS AND DISCUSSION}

\begin{figure}
\includegraphics[scale= 0.4]{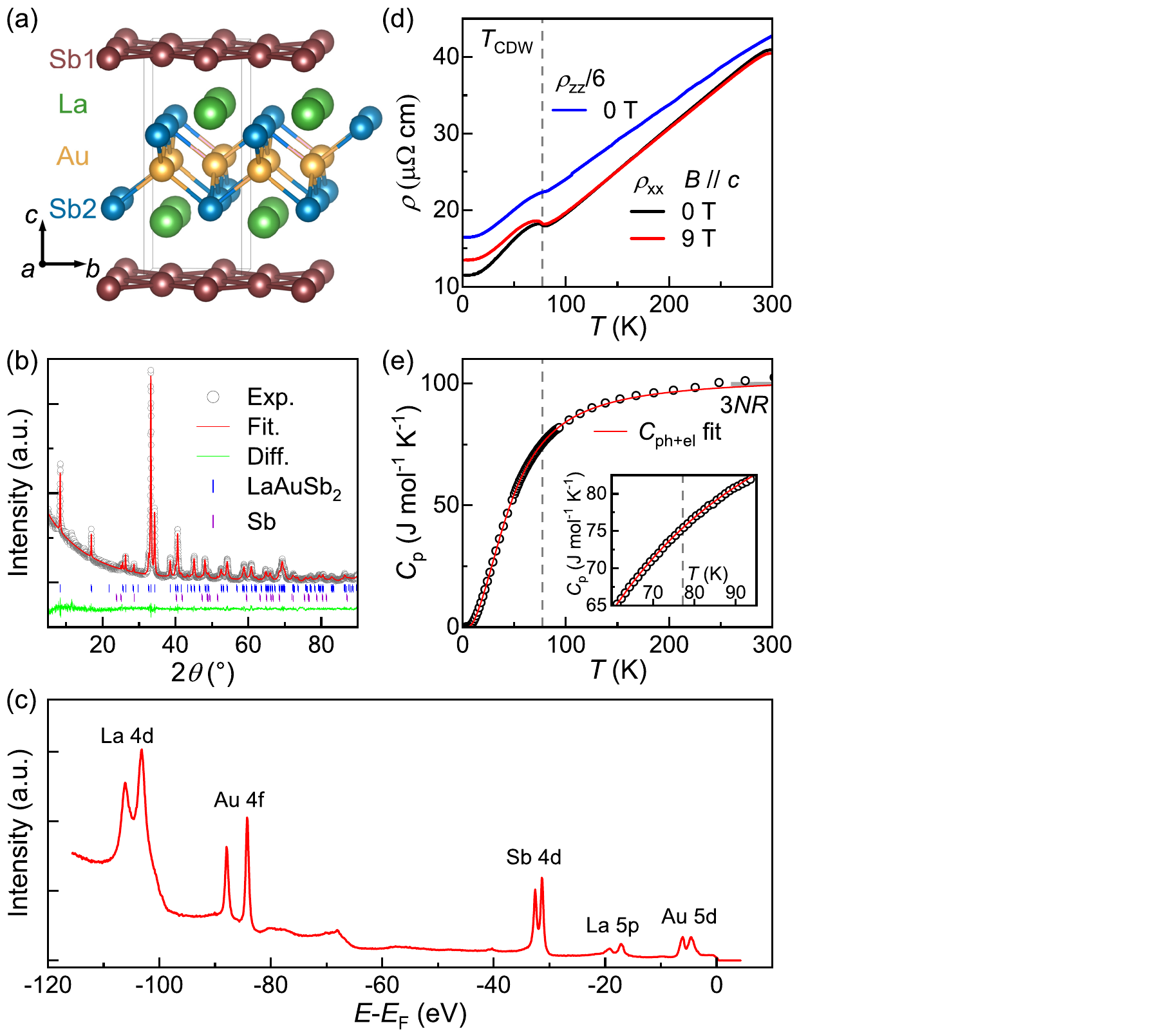}
\caption{ (a) The crystal structure of LaAuSb$_2$. Sb1 and Sb2 denote the Sb located in the Sb-square net and anti-PbO type AuSb layer, respectively (b) Refined powder x-ray diffraction (PXRD) pattern for LaAuSb$_2$. Weak peaks corresponding to a small amount of residual Sb flux were observed in the PXRD pattern. (c) Core-level electronic structure of LaAuSb$_2$, measured using a photon energy of 150 eV. (d) Temperature dependence of in-plane resistivity $\rho_{xx}$ ($j \parallel ab$) at 0 and 9 T and out-of-plane resistivity $\rho_{zz}$ ($j \parallel c$) at 0 T for a LaAuSb$_2$ single crystal, where $j$ is the electric current applied for the resistivity measurements. (e) Temperature dependence of specific heat for LaAuSb$_2$. The inset shows the enlarged view around $T_\mathrm{CDW}$. The red line represents the fit using the Debye-Einstein model.
}
\label{fig1}
\end{figure}

\subsection{XRD, resistivity, and specific heat}

Figure \ref{fig1}(a) shows the crystal structure of LaAuSb$_2$, which is composed of alternatively stacking the Sb$^{1-}$ square net, anti-PbO type [AuSb]$^{2-}$ layer along the $c$ axis, separated by La$^{3+}$ ions. As shown in Fig. \ref{fig1}(b), the powder x-ray diffraction (PXRD) pattern of LaAuSb$_2$ can be well refined using a tetragonal structure (space group $P4/nmm$, No. 129) with lattice parameters $a$ = 4.436(2) {\AA}, $c$ = 10.439(2) {\AA}. The atomic ratio determined by EDX is found to be La: Au: Sb = 25.5: 25.3: 49.2 with a standard deviation of 2\%, suggesting that our crystals are homogeneous. Further, the core level spectra of LaAuSb$_2$ containing all the characteristic peaks of the constituent elements [Fig.~\ref{fig1}(c)] also show that the chemical composition at the surface is highly homogeneous. Xiang \emph{et al.} reported that the physical properties of LaAu$_x$Sb$_2$, especially $T_\mathrm{CDW}$, depend sensitively on Au stoichiometry \cite{xiang_tuning_2020}. We found that the lattice parameters, chemical composition, and $T_{\mathrm{CDW}}$ of our crystal are consistent with LaAu$_{0.94}$Sb$_2$ reported by Xiang \emph{et al.} \cite{xiang_tuning_2020}. However, the apparent EDX-determined composition in our crystals has a lower Au deficiency, possibly due to the lower resolution of our EDX measurements.

 The temperature dependence of the in-plane resistivity $\rho_{xx}(T)$ measured at $B$ = 0 and 9 T with electrical current flowing in an in-plane direction ($j \perp c$) is shown in Fig.  \ref{fig1}(d). $\rho_{xx}(T)$ decreases linearly with temperature decreasing from room temperature to $T_{\mathrm{CDW}}$ = 77 K and then shows a slight increase due to the occurrence of CDW transition. $T_{\mathrm{CDW}}$ is defined as the minimum of the first derivative of resistivity with respect to the temperature [$d{\rho}(T)/dT$], as indicated by the vertical dashed line. The $\rho_{xx}(T)$ below $T_{\mathrm{CDW}}$ = 77 K is significantly enhanced by a magnetic field of 9 T. In other words, the MR value develops with temperature cooling across $T_{\mathrm{CDW}}$, which reaches the maximum value of MR = $[\rho(B) - \rho(0)]/\rho(0) \times 100\%$ = 16\% at 2 K and 9 T. The out-of-plane resistivity $\rho_{zz}(T)$ with the electrical current running along the $c$ axis ($j \parallel c$) is also shown in Fig. \ref{fig1}(d). The behavior of $\rho_{zz}(T)$ curve is similar to that of $\rho_{xx}(T)$, showing a similar residual resistivity ratio (RRR) of $\sim$ 3 with an anisotropy of  $\rho_{zz}(T)/\rho_{xx}(T) \sim$ 6. Note that the resistivity increase induced by CDW transition in $\rho_{zz}(T)$ is weaker than that in $\rho_{xx}(T)$, consistent with the in-plane CDW modulation wave as observed in LaAgSb$_2$ \cite{song_charge-density-wave_2003}. 
 
Specific heat measurements were performed to gain further insight into the CDW transition, and the result is displayed in Fig. \ref{fig1}(e). The value of room-temperature specific heat is in good agreement with the Dulong-Petit limit of 3$NR$, where $N$ = 4 is the atomic number per chemical formula and $R$ = 8.314 \unit{J.mol^{-1}.K^{-1}} is the universal gas constant. As shown in Fig. \ref{fig1}(e), the specific heat can be well described by the Deby-Einstein model \cite{prakash_ferromagnetic_2016}:
\begin{equation}
\begin{aligned} C_\mathrm{el+ph}(T) = & \gamma T+\alpha 9 N R\left(\frac{T}{\theta_\mathrm{D}}\right)^{3} \int_{0}^{\theta_\mathrm{D} / T} \frac{x^{4} e^{x}}{\left(e^{x}-1\right)^{2}} d x \\ &+(1-\alpha) 3 N R \frac{\left(\theta_\mathrm{E} / T\right)^{2} e^{\theta_\mathrm{E} / T}}{\left(e^{\theta_\mathrm{E} / T}-1\right)^{2}} \end{aligned}
\end{equation}
where $\gamma$ is the Sommerfeld coefficient, and $\theta_\mathrm{D}$ and $\theta_\mathrm{E}$ are the Debye and Einstein temperatures, respectively. The coefficients $\alpha$ and $1 - \alpha$ denote the contribution of the Debye and Einstein terms to the phonon heat capacity, respectively. The fitting yelds $\alpha =$ 0.84, $\gamma =$ \SI{5.2}{mJ.mol^{-1}.K^{-2}}, $\theta_\mathrm{D} =$ 207 K, and $\theta_\mathrm{E} =$ 63 K. $\theta_{\mathrm{D}}$ is close to that previously reported in LaAuSb$_2$ ($\theta_\mathrm{D} =$ 196 K) \cite{du_interplay_2020,kuo_characterization_2019}, but significantly lower than that in LaAgSb$_2$ ($\theta_\mathrm{D} =$ 260 K) \cite{akiba_observation_2022}. From the inset of Fig. \ref{fig1}(d), one can see that no anomaly corresponding to the CDW transition can be detected in the specific heat curve, which is reminiscent of the short-ranged CDW in kagome magnet FeGe \cite{teng_discovery_2022}. However, unlike in  FeGe, in the sister compound LaAgSb$_2$, the structural distortions associated with the two CDW transitions are long-ranged, as can be seen in x-ray scattering measurements \cite{song_charge-density-wave_2003}. LaAgSb$_2$ has a higher $T_\mathrm{CDW}$ ($>$ 200 K), but also shows a weak anomaly in specific heat and magnetization and the anomaly in resistivity is even weaker than in LaAuSb$_2$ \cite{lue_weak_2007}. Assuming that LaAuSb$_2$ shares a similar CDW mechanism with LaAgSb$_2$, the CDW in LaAuSb$_2$ very likely has a long-range nature. The CDW order in LaAgSb$_2$ is suggested to be driven by the Fermi surface nesting \cite{shi_observation_2016} resulting in a very large real space periodicity of $\sim$ 17 nm (CDW1, $\tau_1 \sim 0.026 a^*$) but small magnitude of distortions. That, and the absence of CDW gap in ARPES causes only very weak anomalies in physical property measurements \cite{lue_weak_2007}.

\subsection{MR and Hall measurements}

\begin{figure*}
\includegraphics[scale= 0.7 ]{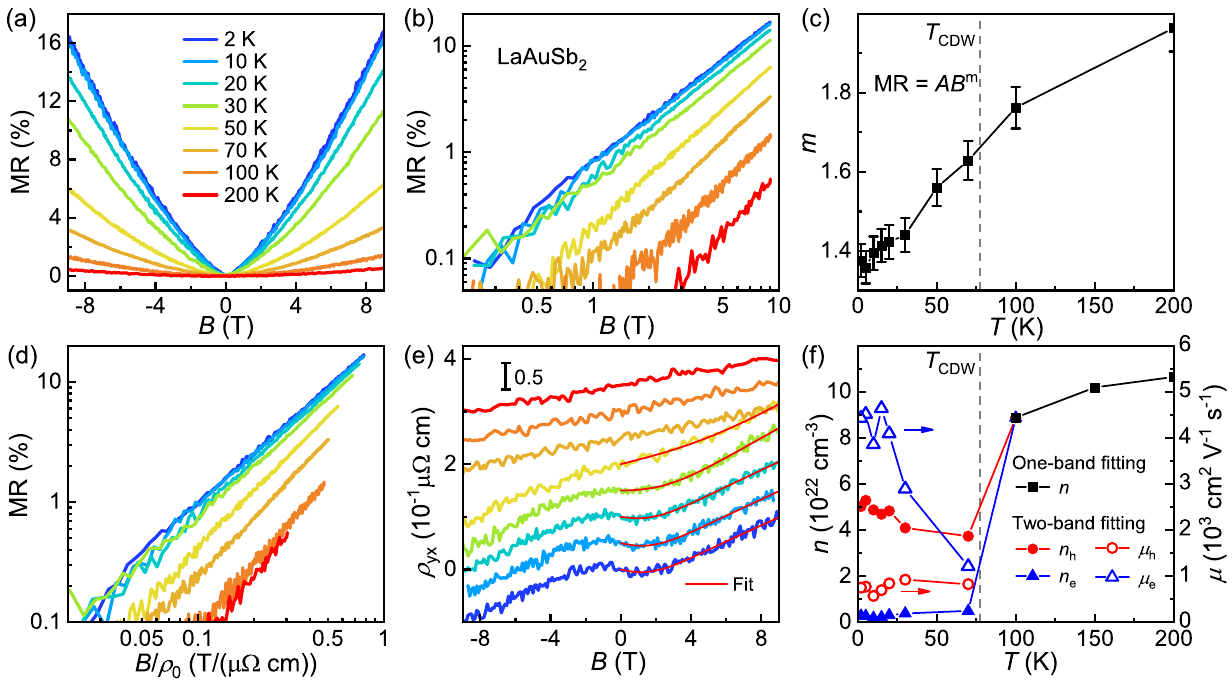}
\caption{(a) Magnetic field dependence of MR at various temperatures with $B \parallel c$. (b) Log-log plot of the MR versus $B$ curves. (c) Temperature dependence of the exponent $n$ in MR = $AB^m$. (d) Kohler plot on log-log scale for LaAuSb$_2$. (e) Hall resistivity $\rho_{yx}$ as a function of the magnetic field at different temperatures with $B \parallel c$. Each subsequent $\rho_{yx}(B)$ curve is shifted upward $0.5 \times 10^{-1} {\mu}{\Omega}$ cm for clarity. The fits using the two-band approach are shown by red lines. (f) Temperature dependence of carrier concentration and mobility, which is estimated by the single-band fitting for $T > T_\mathrm{CDW}$ and the two-band fitting for $T < T_\mathrm{CDW}$. Panels (a), (b), (d), and (e) use the same legend. 
 }
\label{fig2}
\end{figure*}

 Since the behavior of MR could provide valuable information about the underlying electronic structure, MR as a function of the magnetic field was measured at various temperatures and the results are displayed in Fig. \ref{fig2}(a). The maximum MR value is 16\% at 2 K and 9 T,  which decreases monotonously as the temperature increases and becomes negligibly small when $T > T_{\mathrm{CDW}}$, in good agreement with the $\rho(T)$ data [Fig. \ref{fig1}(d)]. To quantify the evolution of the MR($B$) behavior as a function of temperature, we replot the MR($B$) data on a log-log scale in Fig. \ref{fig2}(b) and fit MR($B$) data to a simple power law MR = $AB^m$, where $A$ is a constant, $m$ is the exponent proportional to the slope of the log-log scale plot in Fig. \ref{fig2}(b). The resulting $m$ is plotted against temperature in Fig. \ref{fig2}(c). $m$ increases almost linearly from 1.35 at 2 K to 2 at 200 K without significant anomaly at $T_{\mathrm{CDW}}$. 1.35 $\leq x \leq$ 2 indicates the possible presence of Dirac linear MR together with parabolic MR of topologically trivial states. In a conventional metal, MR($B$) is expected to be parabolic at low fields and it should saturate at high fields, unless perfectly $n–p$ compensated \cite{pippard_magnetoresistance_1989,Pletikosic2014}. Linear MR was often observed in topological semimetals, which was explained in the context of quantum transport. However, the quantum explanation for linear MR often receives criticism because the classical transport, such as inhomogeneity and open Fermi surface \cite{stroud_magnetoresistance_1979,hu_classical_2008}, could also provide a plausible explanation. Although linear MR is not smoking gun proof, we point out that it can still be taken as a signature of the Dirac fermion, whereas the more rigorous confirmation would require further evidence from the band structure. Since the presence of Dirac fermion is indeed indicated by DFT calculations, ARPES, and quantum oscillation measurements (see Sections C--F), the linear MR in LaAuSb$_2$ can be attributed to the quantum transport associated with the Dirac fermion. 

\begin{figure*}[htp!]
\includegraphics[scale= 0.65]{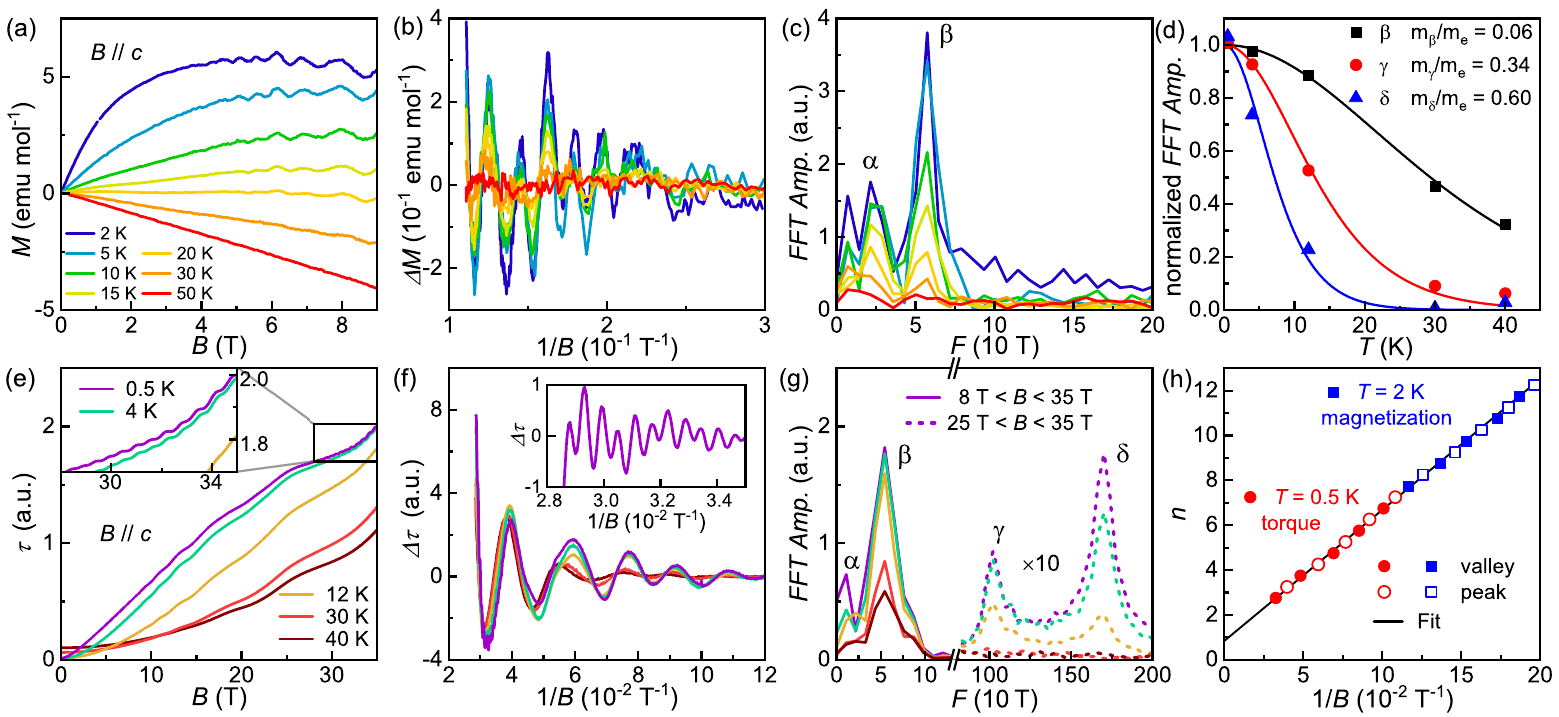}
\caption{(a) Isothermal out-of-plane magnetization ($M(B)$) for LaAuSb$_2$ at various temperatures from 2 K to 50 K. (b) The oscillatory component of magnetization, ${\Delta}M$. (c) The FFT spectra for the oscillatory component ( ${\Delta}M$) for $B \parallel c$. (d) Temperature dependence of the FFT amplitude of the main frequencies, which is obtained from the FFT spectra for $\Delta{\tau}$ in (g). Solid lines represent the fits of the LK formula. (e) The field dependence of magnetic torque $\tau$ for LaAu$_{0.94}$Sb$_2$ at different temperatures. The inset is the enlarged view of the high-field part of $\tau$ as indicated by the black rectangle, where the high-frequency oscillation can be visualized. (f) The oscillatory component of $\tau$ for $B \parallel c$. The inset displays a high-frequency oscillatory component obtained by filtering out low-frequency oscillations. (g) FFT spectra for the dHvA oscillation in the magnetic field range of 8 T $< B <$ 35 T. The dashed lines represent the FFT spectra for the dHvA oscillation in the magnetic field range of 25 T $< B <$ 35 T, which are magnified 10 times in magnitude for clarity. (h) LL index fan diagram for oscillation component ${\Delta}M$ at 2 K and ${\Delta}\tau$ at 0.5 K, where the Berry phase extracted from $\Delta{M}$ is in good agreement with $\Delta{\tau}$. The black line represents a linear fit to all LL indices, which yields an intercept of $n_0 =$ 0.9.}
\label{fig3}
\end{figure*}

To further explore the underlying electronic structure associated with MR($B$), we analyze the MR by Kohler's rule, which dictates that the MR in a conventional metal obeys a scaling behavior of MR = $f(B/{\rho_0})$ \cite{kohler_zur_1938}, where $\rho_0$ is the zero-field resistivity at a given temperature. As shown in Fig. \ref{fig2}(d), instead of collapsing onto a single curve as expected by the Kohler rule, MR vs $B/{\rho_0}$ curves can be divided into three groups, i.e., the low-temperature part ($T \leq$ 30 K), the temperatures around $T_{\mathrm{CDW}}$ (50 and 70 K), and the high-temperature part ($T \geq$ 100 K), demonstrating the violation of Kohler's rule. Kohler's rule is expected to be valid in metal with a single type of charge carrier and a single scattering time, which will be violated in materials with multiple scattering times or phase transitions. Thus, the violation of Kohler's rule in LaAuSb$_2$ is expected and can be understood in terms of CDW transition, i.e., carrier types and densities and scattering times will be dramatically altered by the formation of CDW at temperatures around $T_{\mathrm{CDW}}$, while exhibiting a weak temperature dependency at temperatures $T < T_{\mathrm{CDW}}/2$ and $T > T_{\mathrm{CDW}}$ \cite{mckenzie_violation_1998}, resulting in three different scaling regimes for the MR vs $B/{\rho_0}$ curves. Separate Kohler scaling behavior has also been observed in other CDW systems such as TiSe$_2$ and VSe$_2$ \cite{knowles_fermi_2020,noto_temperature_1980}. In addition, due to the existence of multiple Fermi pockets, as evidenced by the multi-frequencies in quantum oscillations [Fig. \ref{fig3}], nonlinear Hall resistivity [Fig. \ref{fig2}], and measured (in ARPES) and calculated Fermi surface [Figs. \ref{fig5} and \ref{fig6}], the violation of Kohler's rule in LaAuSb$_2$ could also be interpreted in terms of the multi-band scenario.
 
Hall resistivity $\rho_{yx}(B)$ curves are shown in Fig. \ref{fig2}(e). $\rho_{yx}(B)$ exhibits a nearly linear behavior with a slight slope change around $B$ = 0 T for $T \leq$ 70 K, suggesting a multiband behavior with the dominant hole-type carrier. As the temperature increases, the slope of $\rho_{yx}(B)$ varies slightly in the temperature range of 2 $\sim$ 70 K. Meanwhile, the slope change around $B$ = 0 gradually fades away and becomes invisible when $T \geq$ 70 K. The slope of $\rho_{yx}(B)$ exhibits a sudden change as the temperature warms across $T_{\mathrm{CDW}}$, which shows a weak doping dependency with a further increase in temperature. The high-temperature linear $\rho_{yx}(B)$ curves were fitted using a standard single-band model: $d\rho_{yx}(B)/dB = R_{\mathrm{H}} = 1/{ne}$, where $n$ is the carrier concentration and $e$ is the elementary charge. The low-temperature nonlinear $\rho_{yx}(B)$ curves were analysed within a two-band approach \cite{smith_semiconductors_1978}: 
\begin{equation}
\rho_{x y}=\frac{B}{e} \frac{\left(n_{h} \mu_{h}^{2}-n_{e} \mu_{e}^{2}\right)+\left(n_{h}-n_{e}\right)\left(\mu_{h} \mu_{e}\right)^{2} B^{2}}{\left(n_{h} \mu_{h}+n_{e} \mu_{e}\right)^{2}+\left(n_{h}-n_{e}\right)^{2}\left(\mu_{h} \mu_{e}\right)^{2} B^{2}}
\end{equation}
where $n_h$($n_e$) and $\mu_h$($\mu_e$) correspond to the absolute value of the hole (electron) concentration and the hole (electron) mobility, respectively. The fitting results are plotted in Fig. \ref{fig2}(f). The carriers at 200 K are hole-type with the concentration $n \sim$ \SI{1.1e23}{cm^{-3}}, which is significantly suppressed by the formation of CDW order. The transport properties below $T_\mathrm{CDW}$ are dominated by the high concentrations of hole-type carriers with a low mobility and the low concentrations of electron-type carriers with a significantly higher mobility. This result is consistent with the calculated band structure, where the Fermi surface is composed of large hole pockets with parabolic bands [Figs. \ref{fig6}(g) and \ref{fig6}(h)] and relatively small electron pockets with linearly dispersing bands [Figs. \ref{fig6}(i) and \ref{fig6}(j)].

\subsection{Temperature dependence of the dHvA oscillations}

Quantum oscillation measurements could provide direct information on the electronic structure, and serve as one of the most powerful tools during the study of topological semimetals. dHvA oscillation signals in this paper were probed by two strategies, i.e., the magnetization measurements up to 9 T [Fig. \ref{fig3}(a)] and magnetic torque measurements up to 35 T [Fig. \ref{fig3}(e)]. In Fig. \ref{fig3}(a), we present the isothermal out-of-plane ($B \parallel c$) magnetization at various temperatures for a LaAuSb$_2$ single crystal, where clear dHvA oscillations superimposed on a paramagnetic/diamagnetic background can be recognized at $B >$ 4 T. The oscillatory components ${\Delta}M$ extracted by subtracting a smooth background is plotted against 1/$B$ in Fig. \ref{fig3}(b). Note that beat patterns can be discerned on the ${\Delta}M$ vs 1/$B$ curves, indicating that oscillatory components contain multiple frequencies. From the fast Fourier transform (FFT) analyses of the oscillatory components ${\Delta}M$, two frequencies of $F_\alpha =$ 21.5 T and $F_\beta =$ 57.4 T can be derived [Fig. \ref{fig3}(c)]. 

To probe the possible high-frequency oscillation, magnetic torque measurements up to 35 T were performed on a LaAuSb$_2$ single crystal. The obtained magnetic torque $\tau$ and corresponding oscillatory components $\Delta{\tau}$ data are plotted in Figs. \ref{fig3}(e) and \ref{fig3}(f), respectively, showing a low-frequency oscillation as that in $\Delta{M}$ at first glance and no signature of Zeeman splitting can be identified. Upon a close inspection of the high-field part of magnetic torque curves as shown in the inset of Fig. \ref{fig3}(e), high-frequency oscillation with a weak oscillation amplitude can be discerned.  FFT spectra obtained by the FFT analysis in the field range of 8 T $< B <$ 35 T are shown in Fig. \ref{fig3} (g). The main frequency $F_{\beta}$ = 54.7 T is in good agreement with the magnetization measurements. However, the high-frequency peaks are difficult to recognize in the FFT spectra for 8 T $< B <$ 35 T due to the large difference in the oscillation amplitudes. To get a better visualization of high-frequency FFT spectra, FFT analysis was also performed on the oscillatory components $\Delta{\tau}$ in the field range of 25 T $< B <$ 35 T where the low-frequency oscillation can be subtracted as the background [inset in Fig. \ref{fig3}(f)], the obtained FFT spectra are magnified tenfold in amplitude and displayed as dash lines in Fig. \ref{fig3}(g). Figure \ref{fig3}(g) shows four frequencies: $F_{\alpha}$ = 10.9 T, $F_{\beta}$ = 54.7 T, $F_{\gamma}$ = 1021.4 T, and $F_{\delta}$ = 1702.3 T. The oscillation frequency is directly related to the cross-sectional area of the Fermi surface normal to the magnetic field $A_{\mathrm{F}}$ through the Onsager relation \cite{shoenberg_1984}, $F=\left(\frac{\Phi_0}{2\pi^2}\right)A_{\mathrm{F}}$, where $\Phi_0$ is the flux quantum. Then, $A_{\mathrm{F}}$ for the Fermi pockets associated with $F_{\alpha}$, $F_{\beta}$, $F_{\gamma}$, and $F_{\delta}$ are estimated to be 
$A_{\mathrm{F,\alpha}}$ = \SI{0.10}{nm^{-2}}, $A_{\mathrm{F,\beta}}$ = \SI{0.51}{nm^{-2}}, $A_{\mathrm{F,\gamma}}$ = \SI{9.49}{nm^{-2}}, and $A_{\mathrm{F,\delta}}$ = \SI{15.82}{nm^{-2}}. These $A_{\mathrm{F}}$s correspond to 0.05\%, 0.25\%, 4.7\%, and 7.9\% of the total area of the Brillouin zone in the $ab$ plane, suggesting that dHvA oscillations arise from small Fermi pockets.

According to the Lifshitz-Kosevich formula \cite{shoenberg_1984}, the effective mass $m = m^{*}m_e$ can be obtained by fitting the temperature dependence of the oscillation amplitude to the thermal damping factor $R_\mathrm{T}$, which is defined as:
\begin{equation}
R_{\mathrm{T}}=\frac{\alpha T m^{*} / \overline{B}}{\sinh \left(\alpha T m^{*} / \overline{B}\right)} \\
\end{equation} 
where $\alpha=2 \pi^2 k_{\mathrm{B}} m_e / e \hbar=14.69 \mathrm{~T} / \mathrm{K}$, 1/$\overline{B}$ is the average inverse field when performing FFT analysis. As shown in Fig. \ref{fig3}(c), the fittings yield $m_{\beta} = 0.06 m_e$ for Fermi surface $\beta$, which is comparable to that in (Ca, Sr, Ba, Yb)MnSb$_2$ \cite{he_quasi-two-dimensional_2017,liu_magnetic_2017,huang_nontrivial_2017,wang_quantum_2018} but significantly smaller than $0.16 \sim 0.42 m_e$  in LaAgSb$_2$  \cite{myers_haasvan_1999}. Fermi pockets associated with $F_{\gamma}$ and $F_{\delta}$ have larger effective masses. It is well known that effective mass is related to the electronic band curvature $m^*=\hbar^2 /\left[\partial^2 E\left(k\right) / {\partial}k^2\right]$, and hence, small effective masses in LaAuSb$_2$ could serve as evidence for the presence of Dirac linear dispersion.  

In addition to the small Fermi pocket and light effective mass, the Berry phase is another piece of evidence for the Dirac dispersion that could be given by quantum oscillation measurements. A nontrivial $\pi$ Berry phase is expected for the relativistic Dirac fermions with linear dispersion. According to the Lifshitz-Onsager quantization rule, Landau level (LL) index $n$ is related to $1/B$ by $A_{\mathrm{F}}(\hbar/eB) = 2\pi(n + \gamma - \delta)$, from which the Landau level index fan diagram can be constructed. $\gamma - \delta$ is the phase factor for oscillation, where $\gamma = \frac{1}{2} - \frac{\phi_{\mathrm{B}}}{2\pi}$ and $\phi_{\mathrm{B}}$ is the Berry phase, $\delta$ is a phase offset related to the dimensionality of the Fermi surface and takes values of 0 and $\pm$1/8, respectively, for 2D and 3D cases. Berry phase $\phi_{\mathrm{B}}$ is zero for conventional metals while $\pm\pi$ for Dirac fermions. During the analysis of the Berry phase, the minimum of the DOS($E_{\mathrm{F}}$) should be assigned with integer LL indices $n$, that is, the minimum of the Gibbs thermodynamic potential $\Omega$ corresponds to the minimum of DOS($E_{\mathrm{F}}$). The magnetization $M$ is equal to the derivative of the Gibbs thermodynamic potential at constant temperature and chemical potential, i.e., $M = \frac{{\partial}{\Omega}}{{\partial}B}$. Therefore, there should be a $\frac{\pi}{4}$ phase difference between $M$ and DOS($E_{\mathrm{F}}$)  \cite{hu_transport_2019}. When the oscillations in magnetization are used to establish a LL fan diagram, the minima of $M$ should be assigned with $n - 1/4$ (where $n$ is an integer). The obtained Landau level index fan phase diagram is plotted in Fig. \ref{fig3}(h).  where a nontrivial Berry phase with $\gamma$ = 0.9 ($\phi_{\mathrm{B}} = 0.8\pi$) can be deduced from the intercept of the linear fit. This provides evidence for the existence of Dirac fermions in LaAuSb$_2$. Note that the Berry phase deviates from the exact $\pi$ value. This deviation might stem from the presence of parabolic bands around the Fermi level or from the phase offset related to the dimensionality of the Fermi surface. In addition, although the oscillation in LaAuSb$_2$ is composed of four frequencies, the single-band scenario with $F_\beta$ can be a good approximation for constructing the LL index fan diagram due to the large difference in the oscillation amplitudes [Fig. \ref{fig3}(a) and \ref{fig3}(e)]. The applicability of this approximation is further verified by the fact that all the points in Fig. \ref{fig3}(h) fall on a straight line with a small deviation (see more details in supplemental Fig. S2 \cite{SM}).

\begin{figure}[htp!]
\includegraphics[scale= 0.41]{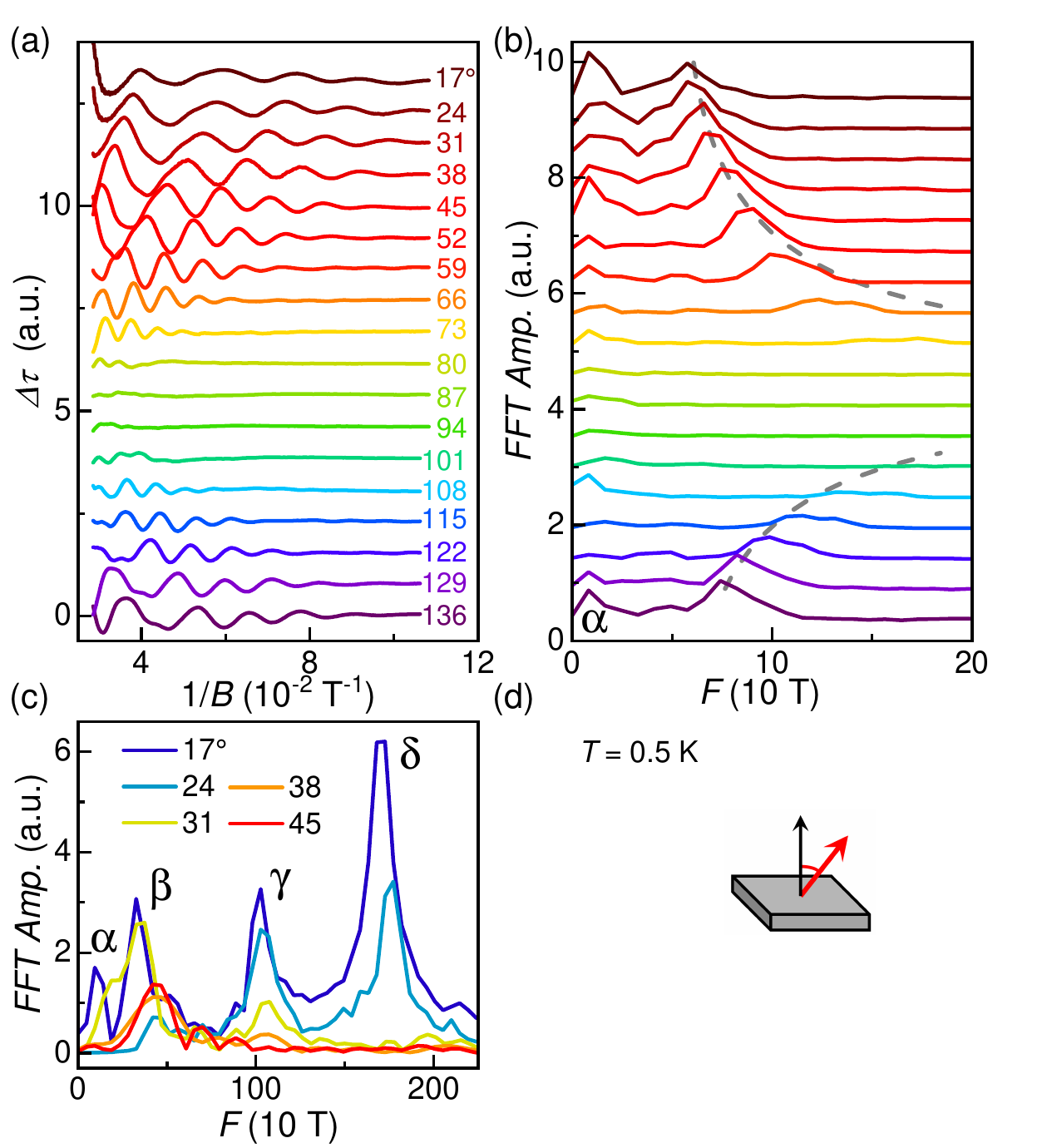}
\caption{(a) dHvA oscillations ($\Delta{\tau}$) at $T$ = 0.5 K for different magnetic field orientations. The FFT spectra of the $\Delta{\tau}$ in the magnetic field range of 8 T $< B <$ 35 T and 25 T $< B <$ 35 T are displayed in (b) and (c), respectively. Data at different field orientations have been shifted for clarity. The dash lines are guides for the eye. (d) The angular dependence of $F_{\alpha}$, the solid lines are the fits to $F = F_{\mathrm{2D}}/{\mathrm{cos}\theta}$. Panels (a-c) use the same legend.
}
\label{fig4}
\end{figure}

\subsection{Angular dependence of quantum oscillations}

To probe the Fermiology and understand the origin of the observed oscillation frequencies, we performed a series of magnetic torque measurements at various magnetic field orientations $\theta$. $\theta$ is the angle between the applied magnetic field and the $c$ axis as illustrated by the measurement setup in the inset of Fig. \ref{fig4}(d). The oscillatory components $\Delta\tau$ measured at various tilt angles $\theta$  are presented in Fig. \ref{fig4}(a). A high-frequency oscillation with a weak amplitude can be recognized for $\theta \leq$ \ang{31} (close to $B \parallel c$). The FFT spectra obtained from the FFT analysis on the field ranges of 8 T $< B <$ 35 T and 25 T $< B <$ 35 T are shown in Figs. \ref{fig4}(b) and \ref{fig4}(c), respectively. $F_{\alpha}$ and $F_{\beta}$ can be discerned in FFT spectra as shown in Fig. \ref{fig4}(b).  As $\theta$ increases from \ang{17} to \ang{136}, $F_{\beta}$ progressively shifts to higher frequencies and decreases in amplitude. The oscillation persists to \ang{80}, becomes invisible in the angle range of \ang{80} $< x <$ \ang{101}, and reappears when $\theta \geq \ang{101}$. The divergence in field orientation around $B \parallel c$ (\ang{80} $< x <$ \ang{101}) is a characteristic of the quasi-2D Fermi surface. For a 2D Fermi surface, the magnetic field projected to the normal axis determines the Landau level quantizations. Thus, the oscillation frequency versus tilt angle curve should follow the inverse of a sinusoidal function. As shown in Fig. \ref{fig4}(d), the $F_{\beta}$ vs $\theta$ curve can be well described by $F = F_{\mathrm{2D}}/\mathrm{cos}(\theta)$, where $F_{\mathrm{2D}}$ is the frequency at $\theta$ = \ang{0}. This indicates that $F_{\beta}$ originates from a quasi-2D cylindrical Fermi surface. From the FFT spectra obtained by the FFT analysis on field range 8 T $< B <$ 35 T, one can see that $F_{\gamma}$ and $F_{\delta}$ also show a quasi-2D character because with the rotation of magnetic field from the $c$ axis, both $F_{\gamma}$ and $F_{\delta}$ decrease in amplitude and become invisible when $\theta > \ang{31}$. Since $F_{\gamma}$ persists to a higher angle, Fermi pocket associated with $F_{\gamma}$ is more 3D than $F_{\delta}$.

\subsection{ARPES}

\begin{figure*}[htp!]
\includegraphics[scale= .8]{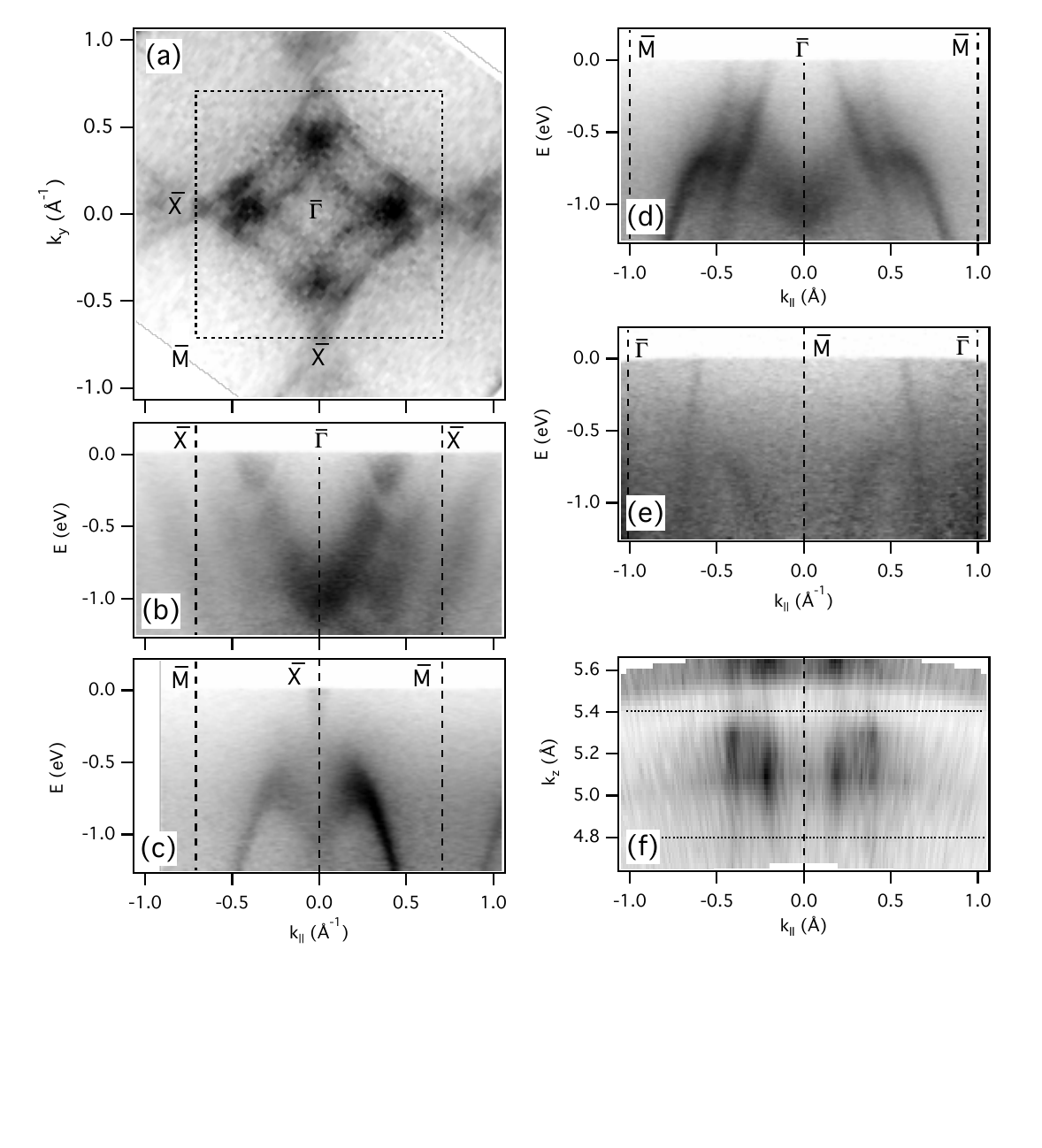}
\caption{Electronic structure of LaAuSb$_2$ from ARPES. (a) The Fermi surface. (b--e) ARPES spectra along the high-symmetry lines of the surface Brillouin zone, as indicated. Spectra in (a--e) were taken using a photon energy of 95 eV, near the $\pi/c$ plane of the 3D BZ at $T$ = 15 K. (f) The Fermi surface in the out-of-plane direction, $k_z$, at $k_y=0$, recorded at photon energies in the range from 76 to 120 eV}
\label{fig5}
\end{figure*}

\begin{figure*}[htp!]
\includegraphics[scale= 0.6]{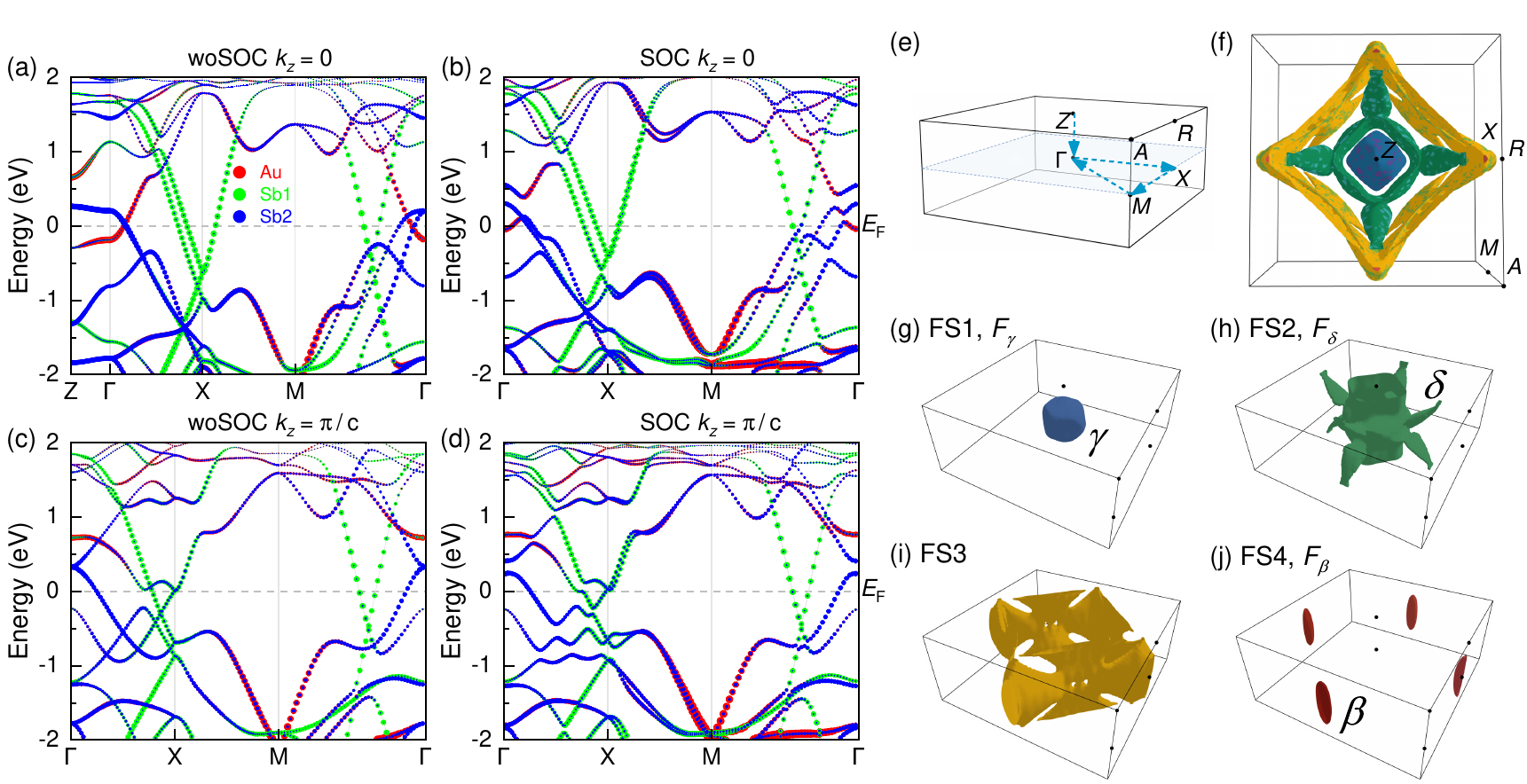}
\caption{Calculated band structures of LaAuSb$_2$ without (a) and with (b) spin-orbit coupling at $k_z$ = 0. Calculated band structures of LaAuSb$_2$ without (c) and with (d) spin-orbit coupling at $k_z$ = $\pi/c$. Red, green, and blue circles represent the contribution from Au, Sb1, and Sb2 orbitals. The area of the circles represents the relative weight of the orbitals. (e) Three-dimensional (3D) Brillouin zone of LaAuSb$_2$. (f) 3D Fermi surface of LaAuSb$_2$ at $E = -0.01$ eV, which consists of four Fermi pockets plotted separately in (g--j) for clarity.}
\label{fig6}
\end{figure*}

ARPES measurements were performed to directly probe the electronic structure of LaAuSb$_2$, as shown in Fig. \ref{fig5}. Panel (a) represents the Fermi surface, while panels ~(b-e) show the ARPES spectra along the high-symmetry directions of the surface Brillouin zone (SBZ). Panel (f) shows the out-of-plane Fermi surface, measured along the $k_x=k_y$ plane. The spectra from the $\bar{\Gamma}-\bar{M}$ line [panel (d)] show at least two linearly-dispersing bands crossing the Fermi level and forming two diamond-shaped Fermi surfaces (FSs) centered at $\bar{\Gamma}$ [Fig.~\ref{fig5}(a)]. The inner states form a diamond-like hole pocket. At first sight, it would seem that the outer bands enclose a large diamond-shaped hole FS. However, this simple picture is deceiving, as demonstrated in panel (e), which shows the same momentum line but sampled in the 2$^{nd}$ surface BZ. Surprisingly, the outer states now display the electron-like dispersion! The explanation for this is in the fact that the large diamonds are actually formed by extremely anisotropic Dirac cones \cite {bosak_evidence_2021}. However, due to the chemical potential shift relative to the LaAgSb$_2$ case, the Dirac point is much closer to the Fermi level and the contour at that energy looks essentially like a Dirac nodal line. One side of the Dirac-like dispersion along the $\bar{\Gamma}-\bar{M}$ is clearly visible in the first BZ, while the other is suppressed in the first, but clearly visible in the second, indicating a strong dependence on ARPES matrix elements, similar to the report by Shi \emph{et al.,} on LaAgSb$_2$  \cite{shi_observation_2016}.
The large diamond-like contours touch at the zone boundaries ($\bar{X}$ points), forming small electron pockets  [Figs.~\ref{fig5}(a--c)]. Additionally, we observe high-intensity patches in between $\bar{\Gamma}$ and $\bar{X}$ points [Fig.~\ref{fig5}(b)], similar to what has been observed in LaAgSb$_2$ \cite{rosmus_electronic_2022}.

Overall, the Fermi surface is similar to that measured in LaAgSb$_2$ \cite{shi_observation_2016,rosmus_electronic_2022} and in good agreement with the DFT calculations shown in Fig. \ref{fig6} \cite{hase_electronic_2014}. 
The estimated area of the inner diamond-like hole pocket is $\sim9$\% of the area of the SBZ. This pocket could then accommodate either $\gamma$ or $\delta$ orbits observed in dHvA, or both, considering its measurable warping in the $k_z$ direction [Fig.\ref{fig5}(f)]. The small electron pockets at $\bar{X}$ points amount to 0.5\%. As discussed in the next section, these pockets could be responsible for the oscillation frequency $F_\beta$, which is the main transport evidence for the existence of Dirac states in LaAuSb$_2$. The situation with the big diamond-like contour is less obvious. The area of the whole diamond is $\sim38\%$ of the area of the SBZ. To observe these orbits in dHvA measurements, much larger fields would be required. If, on the other hand, the observed contour comes from an extremely narrow, anisotropic, and almost "neutral" ($E_D\approx0$) Dirac cone that could, in principle, be an origin of the small $\alpha$ orbit. We note that, in the present case, the ARPES experimental resolution was insufficient to observe such small pockets.
The estimated Fermi velocities of the inner and outer linear bands along $\bar{\Gamma}-\bar{M}$ that form the diamond-shaped FSs are very high: 4.5 and 8.5 eV{\AA}, respectively, nearly two times higher than that observed in LaAgSb$_2$, consistent with the smaller effective mass estimated from the dHvA oscillations \cite{shi_observation_2016}. 

We note that the sides of the smaller and the larger diamond-like Fermi contours are parallel to each other, providing good nesting conditions. If this is the relevant mechanism, the CDW period would have to be around 20 {\AA}, if the modulation is in the crystalographic $a$ direction, or $\sim30$ {\AA}, in the more favorable diagonal direction. If, on the other hand, the relevant nesting is within the slightly split sides of the larger diamond, as suggested by Shi \textit{et al.} \cite{shi_observation_2016}, the period of charge modulation should be much longer than in LaAgSb$_2$, possibly several hundreds of \AA, due to the different filling of these elongated Dirac cones. The period of CDW modulations in LaAuSb$_2$ remains to be checked, as detailed structural studies are still lacking. In both nesting scenarios, the nested FS segments are expected to be gapped in the CDW state. However, our ARPES spectra do not display any detectable gaps, similar to some other quasi-2D materials displaying CDW phase \cite{2H_NbSe2,Valla2006}.

\subsection{Electronic structure calculations}

In order to better understand the magnetotransport and ARPES data, we performed comprehensive calculations of the electronic structure on LaAuSb$_2$. To emphasize the effects of SOC and $k_z$ dependency, the calculated band structures are plotted into four panels as shown in Figs. \ref{fig6}(a--d), which are consistent with previous reports \cite{hase_electronic_2014}, and share common features with square-net-based topological semimetals, such as LaCuSb$_2$\cite{chamorro_dirac_2019,akiba_phonon-mediated_2023}, LaAgSb$_2$\cite{rosmus_electronic_2022}, CaMnBi$_2$\cite{feng_strong_2015}, YbMnBi$_2$\cite{borisenko_time-reversal_2019}, BaZnBi$_2$\cite{zhao_quantum_2018,thirupathaiah_absence_2019}, and ZrSiS\cite{schoop_dirac_2016}. 

From  Figs. \ref{fig6}(a--d), one can see that four bands crossing the Fermi level ($E_\mathrm{F}$), forming a Fermi surface with four Fermi pockets [Fig. \ref{fig6}(f)]. The Fermi pockets are labeled as FS1--4 and plotted separately into Figs. \ref{fig6}(g--j) for clarity.

As shown in Figs. \ref{fig6}(a--d), the bands arising from the orbitals of AuSb layer crossing $E_\mathrm{F}$ around the $\Gamma$ point, forming the Fermi pockets of FS1 and FS2. As can be seen in Figs. \ref{fig6}(g) and \ref{fig6}(h), FS1 is a 3D Fermi pocket with a nearly square-shaped cross section while FS2 is a pillar-like 2D Fermi pocket with eight branches along the $\Gamma - X$ direction (four upward and four downward).

The linear bands crossing $E_\mathrm{F}$ at the midpoint of the $\Gamma - M$ line originate from the 5$p_x$/5$p_y$ orbitals of Sb1 located in the square net, intersect at a Dirac point at $\sim$ 0.35 eV below $E_\mathrm{F}$. Note that there is a small gap (55 meV) at the Dirac point, which increases slightly to 71 meV when SOC is taken into account. These linear bands form a large diamond-shaped electron Fermi pocket FS3 as shown in Fig. \ref{fig6}(i). The linear bands and the diamond-like Fermi pocket are the intrinsic features for the square-net structure, which has been experimentally observed in LaAgSb$_2$\cite{rosmus_electronic_2022}, CaMnBi$_2$\cite{feng_strong_2015}, YbMnBi$_2$\cite{borisenko_time-reversal_2019}, BaZnBi$_2$\cite{zhao_quantum_2018}, and ZrSiS\cite{schoop_dirac_2016}.

The linear bands crossing $E_\mathrm{F}$ at the $X$ point also arise from the Sb-square net, intersecting at two Dirac points: one at $E = -$ 0.6 eV with a small gap of 70 meV and another at $E = -$ 0.4 eV with a large gap of 350 meV. These gaps at Driac points are also slightly increased by SOC effects. Notice that these linear bands strongly depend on $k_z$, indicating a three-dimensional (3D) character. According to their $k_z$ dependency, the inner linear bands should account for the ellipsoidal shape electron Fermi pocket FS4 centered at the $X$ point [Fig. \ref{fig6}(j)], while the outer linear bands can be assigned to the large diamond-like Fermi pocket FS [Fig. \ref{fig6}(i)].

We first compare the calculated electronic structures with the ARPES data. As shown in Fig. S4 in Supplemental Material \cite{SM}, the Fermi surfaces exhibit a strong Fermi energy dependency: as the Fermi energy decreases, the pillar-shaped FS2 with eight branches becomes fatter while the diamond-shaped FS3 shrinks and eventually splits into small sheets. From Figs. S4 and S5 in the Supplemental Material \cite{SM}, we found that the Fermi surfaces at $E = -0.01$ eV match well with the Fermi surface measured by ARPES [Fig. \ref{fig5} (a)].  This Fermi surface could also provide a good explanation for the dHvA oscillation measurements. Through the comparison with the cross-sectional areas of the calculated Fermi pockets with that inferred from the oscillation frequencies [Figs. S4 and S5 in the Supplemental Material \cite{SM}], one can see that $F_\beta$, $F_\gamma$, and $F_\delta$ can be assigned to the ellipsoidal shape FS4 centered at the $X$ point, the inner 3D square-shaped FS1 around $\Gamma$ point, and the outer pillar-shaped FS2 around $\Gamma$ point, respectively. These assignments are further verified by the good agreement of band dispersions inferred from the effective masses, ARPES, and DFT calculations.

\subsection{Discussion}

Although our magnetotransport, ARPES, and DFT studies provide strong evidence for LaAuSb$_2$ to be a Dirac semimetal, its transport properties deviate from prototypical Dirac semimetals, such as Cd$_3$As$_2$, which exhibits a low carrier concentration, ultrahigh mobility, and giant linear MR \cite{liang_ultrahigh_2015}. The differences originate from different electronic structures: Fermi surface of Cd$_3$As$_2$ is composed of a pair of point-like Fermi pockets with a sharp linear dispersion (the Fermi velocity is determined as $V_x \sim$ 8 eV\AA) \cite{liu_stable_2014,neupane_observation_2014}; LaAuSb$_2$ has a more complex Fermi surface with four Fermi pockets, not all of them originating from the Dirac bands. As shown in the ARPES section, Fermi velocities in certain parts of Fermi pockets of LaAuSb$_2$ also have values comparable to that in Cd$_3$As$_2$, but the average Fermi velocity is smaller due to the anisotropy and the presence of Fermi pockets with small Fermi velocities. Thus, in LaAuSb$_2$, only a portion of the carriers originate from Dirac bands, and even those are highly anisotropic (elongated), in contrast to highly isotropic 3D-Dirac bands that exist in Cd$_3$As$_2$, without conventional bands spoiling the Dirac-like character of carriers.      

Recently, the Kagome lattice has attracted great attention due to the observation of CDW, superconductivity, and Dirac fermion. Prime examples include CDW, electronic nematicity, and superconductivity in CsV$_3$Sb$_5$ \cite{ortiz_cs_2020,nie_CDW-driven_2022}, CDW in ScV$_6$Sn$_6$ \cite{arachchige_charge_2022}, and CDW in FeGe antiferromagnet \cite{teng_discovery_2022}. The emergence of these interesting electronic orders in the kagome lattice comes from the unique lattice geometry of the kagome lattice, which will naturally lead to the flat band, the topological Dirac crossing at the $K$ point, and a pair of van Hove singularities (vHSs) at $M$ points. For example, it is suggested that CDW in CsV$_3$Sb$_5$ is triggered by vHSs \cite{tan_charge_2021}. From the present study, we see that LaAuSb$_2$ could also simultaneously host Dirac fermions, CDW, and superconductivity, making the square net an ideal model system for investigating these electronic orders. 

\section{CONCLUSION}
In conclusion, we have studied the physical properties and underlying electronic structure of LaAuSb$_2$ single crystals using magnetotransport and ARPES measurements and DFT calculations. Resistivity and Hall measurements reveal the occurrence of the CDW transition at 77 K. The quantum oscillations and ARPES data reveal quasi-2D Fermi surfaces with nearly massless Dirac fermions, in good agreement with the DFT calculations. Our results indicate that LaAuSb$_2$ hosts CDW and Dirac fermions at the same time, suggesting that the Sb square net could be a building block for designing topological materials with various electronic orders.

\begin{acknowledgments}
The work in Chongqing was financially supported in part by the National Natural Science Foundation of China (Grant No. 12004056, No. 12104072), the Chongqing Basic Research and Frontier Technology, China (Grants No. cstc2021jcyj-msxmX0661, No. cstc2021jcyj-msxmX0640), Fundamental Research Funds for the Central Universities, China (Grants No. 2022CDJXY-002, No. 2023CDJXY-048). Work at BNL was supported by US DOE, Office of Science, and Office of Basic Energy Sciences under contract DE-SC0012704. Work at the National High Magnetic Field Laboratory is supported by National Science Foundation Cooperative Agreement No. DMR-1644779 and the State of Florida. T.V. acknowledges the support from the Red guipuzcoana de Ciencia, Tecnología e Innovación – Gipuzkoa NEXT 2023 from the Gipuzkoa Provincial Council. The authors thank Yan Liu and Guiwen Wang at the Analytical and Testing Center of Chongqing University and Elio Vescovo and Anil Rajapitamahuni of NSLS II for their assistance with measurements.
\end{acknowledgments}


\begin{thebibliography}{57}%
\makeatletter
\providecommand \@ifxundefined [1]{%
 \@ifx{#1\undefined}
}%
\providecommand \@ifnum [1]{%
 \ifnum #1\expandafter \@firstoftwo
 \else \expandafter \@secondoftwo
 \fi
}%
\providecommand \@ifx [1]{%
 \ifx #1\expandafter \@firstoftwo
 \else \expandafter \@secondoftwo
 \fi
}%
\providecommand \natexlab [1]{#1}%
\providecommand \enquote  [1]{``#1''}%
\providecommand \bibnamefont  [1]{#1}%
\providecommand \bibfnamefont [1]{#1}%
\providecommand \citenamefont [1]{#1}%
\providecommand \href@noop [0]{\@secondoftwo}%
\providecommand \href [0]{\begingroup \@sanitize@url \@href}%
\providecommand \@href[1]{\@@startlink{#1}\@@href}%
\providecommand \@@href[1]{\endgroup#1\@@endlink}%
\providecommand \@sanitize@url [0]{\catcode `\\12\catcode `\$12\catcode
  `\&12\catcode `\#12\catcode `\^12\catcode `\_12\catcode `\%12\relax}%
\providecommand \@@startlink[1]{}%
\providecommand \@@endlink[0]{}%
\providecommand \url  [0]{\begingroup\@sanitize@url \@url }%
\providecommand \@url [1]{\endgroup\@href {#1}{\urlprefix }}%
\providecommand \urlprefix  [0]{URL }%
\providecommand \Eprint [0]{\href }%
\providecommand \doibase [0]{https://doi.org/}%
\providecommand \selectlanguage [0]{\@gobble}%
\providecommand \bibinfo  [0]{\@secondoftwo}%
\providecommand \bibfield  [0]{\@secondoftwo}%
\providecommand \translation [1]{[#1]}%
\providecommand \BibitemOpen [0]{}%
\providecommand \bibitemStop [0]{}%
\providecommand \bibitemNoStop [0]{.\EOS\space}%
\providecommand \EOS [0]{\spacefactor3000\relax}%
\providecommand \BibitemShut  [1]{\csname bibitem#1\endcsname}%
\let\auto@bib@innerbib\@empty
\bibitem [{\citenamefont {Song}\ \emph {et~al.}(2003)\citenamefont {Song},
  \citenamefont {Park}, \citenamefont {Koo}, \citenamefont {Lee}, \citenamefont
  {Rhee}, \citenamefont {Bud’ko}, \citenamefont {Canfield}, \citenamefont
  {Harmon},\ and\ \citenamefont {Goldman}}]{song_charge-density-wave_2003}%
  \BibitemOpen
  \bibfield  {author} {\bibinfo {author} {\bibfnamefont {C.}~\bibnamefont
  {Song}}, \bibinfo {author} {\bibfnamefont {J.}~\bibnamefont {Park}}, \bibinfo
  {author} {\bibfnamefont {J.}~\bibnamefont {Koo}}, \bibinfo {author}
  {\bibfnamefont {K.-B.}\ \bibnamefont {Lee}}, \bibinfo {author} {\bibfnamefont
  {J.~Y.}\ \bibnamefont {Rhee}}, \bibinfo {author} {\bibfnamefont {S.~L.}\
  \bibnamefont {Bud’ko}}, \bibinfo {author} {\bibfnamefont {P.~C.}\
  \bibnamefont {Canfield}}, \bibinfo {author} {\bibfnamefont {B.~N.}\
  \bibnamefont {Harmon}},\ and\ \bibinfo {author} {\bibfnamefont {A.~I.}\
  \bibnamefont {Goldman}},\ }\bibfield  {title} {\bibinfo {title}
  {Charge-density-wave orderings in {LaAgSb$_2$}: {An} x-ray scattering
  study},\ }\href {https://doi.org/10.1103/PhysRevB.68.035113} {\bibfield
  {journal} {\bibinfo  {journal} {Physical Review B}\ }\textbf {\bibinfo
  {volume} {68}},\ \bibinfo {pages} {035113} (\bibinfo {year}
  {2003})}\BibitemShut {NoStop}%
\bibitem [{\citenamefont {Akiba}\ \emph
  {et~al.}(2022{\natexlab{a}})\citenamefont {Akiba}, \citenamefont {Umeshita},\
  and\ \citenamefont {Kobayashi}}]{akiba_observation_2022}%
  \BibitemOpen
  \bibfield  {author} {\bibinfo {author} {\bibfnamefont {K.}~\bibnamefont
  {Akiba}}, \bibinfo {author} {\bibfnamefont {N.}~\bibnamefont {Umeshita}},\
  and\ \bibinfo {author} {\bibfnamefont {T.~C.}\ \bibnamefont {Kobayashi}},\
  }\bibfield  {title} {\bibinfo {title} {Observation of superconductivity and
  its enhancement at the charge density wave critical point in {LaAgSb$_2$}},\
  }\href {https://doi.org/10.1103/PhysRevB.106.L161113} {\bibfield  {journal}
  {\bibinfo  {journal} {Physical Review B}\ }\textbf {\bibinfo {volume}
  {106}},\ \bibinfo {pages} {L161113} (\bibinfo {year} {2022}{\natexlab{a}})},\
  \bibinfo {note} {publisher: American Physical Society}\BibitemShut {NoStop}%
\bibitem [{\citenamefont {Shi}\ \emph {et~al.}(2016)\citenamefont {Shi},
  \citenamefont {Richard}, \citenamefont {Wang}, \citenamefont {Liu},
  \citenamefont {Matt}, \citenamefont {Xu}, \citenamefont {Dhaka},
  \citenamefont {Ristic}, \citenamefont {Qian}, \citenamefont {Yang},
  \citenamefont {Petrovic}, \citenamefont {Shi},\ and\ \citenamefont
  {Ding}}]{shi_observation_2016}%
  \BibitemOpen
  \bibfield  {author} {\bibinfo {author} {\bibfnamefont {X.}~\bibnamefont
  {Shi}}, \bibinfo {author} {\bibfnamefont {P.}~\bibnamefont {Richard}},
  \bibinfo {author} {\bibfnamefont {K.}~\bibnamefont {Wang}}, \bibinfo {author}
  {\bibfnamefont {M.}~\bibnamefont {Liu}}, \bibinfo {author} {\bibfnamefont
  {C.~E.}\ \bibnamefont {Matt}}, \bibinfo {author} {\bibfnamefont
  {N.}~\bibnamefont {Xu}}, \bibinfo {author} {\bibfnamefont {R.~S.}\
  \bibnamefont {Dhaka}}, \bibinfo {author} {\bibfnamefont {Z.}~\bibnamefont
  {Ristic}}, \bibinfo {author} {\bibfnamefont {T.}~\bibnamefont {Qian}},
  \bibinfo {author} {\bibfnamefont {Y.-F.}\ \bibnamefont {Yang}}, \bibinfo
  {author} {\bibfnamefont {C.}~\bibnamefont {Petrovic}}, \bibinfo {author}
  {\bibfnamefont {M.}~\bibnamefont {Shi}},\ and\ \bibinfo {author}
  {\bibfnamefont {H.}~\bibnamefont {Ding}},\ }\bibfield  {title} {\bibinfo
  {title} {Observation of {Dirac}-like band dispersion in {LaAgSb$_2$}},\
  }\href {https://doi.org/10.1103/PhysRevB.93.081105} {\bibfield  {journal}
  {\bibinfo  {journal} {Physical Review B}\ }\textbf {\bibinfo {volume} {93}},\
  \bibinfo {pages} {081105} (\bibinfo {year} {2016})}\BibitemShut {NoStop}%
\bibitem [{\citenamefont {Baek}\ \emph {et~al.}(2022)\citenamefont {Baek},
  \citenamefont {Bud'ko}, \citenamefont {Canfield}, \citenamefont {Borsa},\
  and\ \citenamefont {Suh}}]{baek_nmr_2022}%
  \BibitemOpen
  \bibfield  {author} {\bibinfo {author} {\bibfnamefont {S.-H.}\ \bibnamefont
  {Baek}}, \bibinfo {author} {\bibfnamefont {S.~L.}\ \bibnamefont {Bud'ko}},
  \bibinfo {author} {\bibfnamefont {P.~C.}\ \bibnamefont {Canfield}}, \bibinfo
  {author} {\bibfnamefont {F.}~\bibnamefont {Borsa}},\ and\ \bibinfo {author}
  {\bibfnamefont {B.~J.}\ \bibnamefont {Suh}},\ }\bibfield  {title} {\bibinfo
  {title} {{NMR} evidence for a {Peierls} transition in the layered square-net
  compound {LaAgSb$_2$}},\ }\href {https://doi.org/10.1103/PhysRevB.106.195153}
  {\bibfield  {journal} {\bibinfo  {journal} {Physical Review B}\ }\textbf
  {\bibinfo {volume} {106}},\ \bibinfo {pages} {195153} (\bibinfo {year}
  {2022})},\ \bibinfo {note} {publisher: American Physical Society}\BibitemShut
  {NoStop}%
\bibitem [{\citenamefont {Bosak}\ \emph {et~al.}(2021)\citenamefont {Bosak},
  \citenamefont {Souliou}, \citenamefont {Faugeras}, \citenamefont {Heid},
  \citenamefont {Molas}, \citenamefont {Chen}, \citenamefont {Wang},
  \citenamefont {Potemski},\ and\ \citenamefont
  {Le~Tacon}}]{bosak_evidence_2021}%
  \BibitemOpen
  \bibfield  {author} {\bibinfo {author} {\bibfnamefont {A.}~\bibnamefont
  {Bosak}}, \bibinfo {author} {\bibfnamefont {S.-M.}\ \bibnamefont {Souliou}},
  \bibinfo {author} {\bibfnamefont {C.}~\bibnamefont {Faugeras}}, \bibinfo
  {author} {\bibfnamefont {R.}~\bibnamefont {Heid}}, \bibinfo {author}
  {\bibfnamefont {M.~R.}\ \bibnamefont {Molas}}, \bibinfo {author}
  {\bibfnamefont {R.-Y.}\ \bibnamefont {Chen}}, \bibinfo {author}
  {\bibfnamefont {N.-L.}\ \bibnamefont {Wang}}, \bibinfo {author}
  {\bibfnamefont {M.}~\bibnamefont {Potemski}},\ and\ \bibinfo {author}
  {\bibfnamefont {M.}~\bibnamefont {Le~Tacon}},\ }\bibfield  {title} {\bibinfo
  {title} {Evidence for nesting-driven charge density wave instabilities in the
  quasi-two-dimensional material {LaAgSb$_2$}},\ }\href
  {https://doi.org/10.1103/PhysRevResearch.3.033020} {\bibfield  {journal}
  {\bibinfo  {journal} {Physical Review Research}\ }\textbf {\bibinfo {volume}
  {3}},\ \bibinfo {pages} {033020} (\bibinfo {year} {2021})}\BibitemShut
  {NoStop}%
\bibitem [{\citenamefont {Wang}\ and\ \citenamefont
  {Petrovic}(2012)}]{wang_multiband_2012}%
  \BibitemOpen
  \bibfield  {author} {\bibinfo {author} {\bibfnamefont {K.}~\bibnamefont
  {Wang}}\ and\ \bibinfo {author} {\bibfnamefont {C.}~\bibnamefont
  {Petrovic}},\ }\bibfield  {title} {\bibinfo {title} {Multiband effects and
  possible {Dirac} states in {LaAgSb$_2$}},\ }\href
  {https://doi.org/10.1103/PhysRevB.86.155213} {\bibfield  {journal} {\bibinfo
  {journal} {Physical Review B}\ }\textbf {\bibinfo {volume} {86}},\ \bibinfo
  {pages} {155213} (\bibinfo {year} {2012})}\BibitemShut {NoStop}%
\bibitem [{\citenamefont {Akiba}\ \emph
  {et~al.}(2022{\natexlab{b}})\citenamefont {Akiba}, \citenamefont {Umeshita},\
  and\ \citenamefont {Kobayashi}}]{Akiba_Magnetotransport_2022}%
  \BibitemOpen
  \bibfield  {author} {\bibinfo {author} {\bibfnamefont {K.}~\bibnamefont
  {Akiba}}, \bibinfo {author} {\bibfnamefont {N.}~\bibnamefont {Umeshita}},\
  and\ \bibinfo {author} {\bibfnamefont {T.~C.}\ \bibnamefont {Kobayashi}},\
  }\bibfield  {title} {\bibinfo {title} {Magnetotransport studies of the {Sb}
  square-net compound {${\mathrm{LaAgSb}}_{2}$} under high pressure and
  rotating magnetic fields},\ }\href
  {https://doi.org/10.1103/PhysRevB.105.035108} {\bibfield  {journal} {\bibinfo
   {journal} {Phys. Rev. B}\ }\textbf {\bibinfo {volume} {105}},\ \bibinfo
  {pages} {035108} (\bibinfo {year} {2022}{\natexlab{b}})}\BibitemShut
  {NoStop}%
\bibitem [{\citenamefont {Klemenz}\ \emph {et~al.}(2019)\citenamefont
  {Klemenz}, \citenamefont {Lei},\ and\ \citenamefont
  {Schoop}}]{klemenz_topological_2019}%
  \BibitemOpen
  \bibfield  {author} {\bibinfo {author} {\bibfnamefont {S.}~\bibnamefont
  {Klemenz}}, \bibinfo {author} {\bibfnamefont {S.}~\bibnamefont {Lei}},\ and\
  \bibinfo {author} {\bibfnamefont {L.~M.}\ \bibnamefont {Schoop}},\ }\bibfield
   {title} {\bibinfo {title} {Topological {Semimetals} in {Square}-{Net}
  {Materials}},\ }\href {https://doi.org/10.1146/annurev-matsci-070218-010114}
  {\bibfield  {journal} {\bibinfo  {journal} {Annual Review of Materials
  Research}\ }\textbf {\bibinfo {volume} {49}},\ \bibinfo {pages} {185}
  (\bibinfo {year} {2019})}\BibitemShut {NoStop}%
\bibitem [{\citenamefont {Xia}\ \emph {et~al.}(2023)\citenamefont {Xia},
  \citenamefont {Wang}, \citenamefont {Zhu}, \citenamefont {Zhang},
  \citenamefont {Liu}, \citenamefont {Wu}, \citenamefont {Zhang}, \citenamefont
  {Yang}, \citenamefont {Yang}, \citenamefont {He}, \citenamefont {Chai},
  \citenamefont {Fu}, \citenamefont {Zhou},\ and\ \citenamefont
  {Wang}}]{xia_coupling_2023}%
  \BibitemOpen
  \bibfield  {author} {\bibinfo {author} {\bibfnamefont {Y.}~\bibnamefont
  {Xia}}, \bibinfo {author} {\bibfnamefont {L.}~\bibnamefont {Wang}}, \bibinfo
  {author} {\bibfnamefont {Y.}~\bibnamefont {Zhu}}, \bibinfo {author}
  {\bibfnamefont {L.}~\bibnamefont {Zhang}}, \bibinfo {author} {\bibfnamefont
  {Y.}~\bibnamefont {Liu}}, \bibinfo {author} {\bibfnamefont {X.}~\bibnamefont
  {Wu}}, \bibinfo {author} {\bibfnamefont {L.}~\bibnamefont {Zhang}}, \bibinfo
  {author} {\bibfnamefont {T.}~\bibnamefont {Yang}}, \bibinfo {author}
  {\bibfnamefont {K.}~\bibnamefont {Yang}}, \bibinfo {author} {\bibfnamefont
  {M.}~\bibnamefont {He}}, \bibinfo {author} {\bibfnamefont {Y.}~\bibnamefont
  {Chai}}, \bibinfo {author} {\bibfnamefont {H.}~\bibnamefont {Fu}}, \bibinfo
  {author} {\bibfnamefont {X.}~\bibnamefont {Zhou}},\ and\ \bibinfo {author}
  {\bibfnamefont {A.}~\bibnamefont {Wang}},\ }\bibfield  {title} {\bibinfo
  {title} {Coupling between the spatially separated magnetism and the
  topological band revealed by magnetotransport measurements on
  {EuMn$_{1-x}$Zn$_x$Sb$_2$} (0 $\leq x \leq$ 1)},\ }\href
  {https://doi.org/10.1103/PhysRevB.108.165115} {\bibfield  {journal} {\bibinfo
   {journal} {Physical Review B}\ }\textbf {\bibinfo {volume} {108}},\ \bibinfo
  {pages} {165115} (\bibinfo {year} {2023})}\BibitemShut {NoStop}%
\bibitem [{\citenamefont {Muro}\ \emph {et~al.}(1997)\citenamefont {Muro},
  \citenamefont {Takeda},\ and\ \citenamefont {Ishikawa}}]{muro_magnetic_1997}%
  \BibitemOpen
  \bibfield  {author} {\bibinfo {author} {\bibfnamefont {Y.}~\bibnamefont
  {Muro}}, \bibinfo {author} {\bibfnamefont {N.}~\bibnamefont {Takeda}},\ and\
  \bibinfo {author} {\bibfnamefont {M.}~\bibnamefont {Ishikawa}},\ }\bibfield
  {title} {\bibinfo {title} {Magnetic and transport properties of dense {Kondo}
  systems, {Ce$T$Sb$_2$} ({$T$} = {Ni}, {Cu}, {Pd} and {Ag})},\ }\href
  {https://doi.org/10.1016/S0925-8388(96)03128-3} {\bibfield  {journal}
  {\bibinfo  {journal} {Journal of Alloys and Compounds}\ }\textbf {\bibinfo
  {volume} {257}},\ \bibinfo {pages} {23} (\bibinfo {year} {1997})}\BibitemShut
  {NoStop}%
\bibitem [{\citenamefont {Chamorro}\ \emph {et~al.}(2019)\citenamefont
  {Chamorro}, \citenamefont {Topp}, \citenamefont {Fang}, \citenamefont
  {Winiarski}, \citenamefont {Ast}, \citenamefont {Krivenkov}, \citenamefont
  {Varykhalov}, \citenamefont {Ramshaw}, \citenamefont {Schoop},\ and\
  \citenamefont {McQueen}}]{chamorro_dirac_2019}%
  \BibitemOpen
  \bibfield  {author} {\bibinfo {author} {\bibfnamefont {J.~R.}\ \bibnamefont
  {Chamorro}}, \bibinfo {author} {\bibfnamefont {A.}~\bibnamefont {Topp}},
  \bibinfo {author} {\bibfnamefont {Y.}~\bibnamefont {Fang}}, \bibinfo {author}
  {\bibfnamefont {M.~J.}\ \bibnamefont {Winiarski}}, \bibinfo {author}
  {\bibfnamefont {C.~R.}\ \bibnamefont {Ast}}, \bibinfo {author} {\bibfnamefont
  {M.}~\bibnamefont {Krivenkov}}, \bibinfo {author} {\bibfnamefont
  {A.}~\bibnamefont {Varykhalov}}, \bibinfo {author} {\bibfnamefont {B.~J.}\
  \bibnamefont {Ramshaw}}, \bibinfo {author} {\bibfnamefont {L.~M.}\
  \bibnamefont {Schoop}},\ and\ \bibinfo {author} {\bibfnamefont {T.~M.}\
  \bibnamefont {McQueen}},\ }\bibfield  {title} {\bibinfo {title} {Dirac
  fermions and possible weak antilocalization in {LaCuSb$_2$}},\ }\href
  {https://doi.org/10.1063/1.5124685} {\bibfield  {journal} {\bibinfo
  {journal} {APL Materials}\ }\textbf {\bibinfo {volume} {7}},\ \bibinfo
  {pages} {121108} (\bibinfo {year} {2019})}\BibitemShut {NoStop}%
\bibitem [{\citenamefont {Akiba}\ and\ \citenamefont
  {Kobayashi}(2023)}]{akiba_phonon-mediated_2023}%
  \BibitemOpen
  \bibfield  {author} {\bibinfo {author} {\bibfnamefont {K.}~\bibnamefont
  {Akiba}}\ and\ \bibinfo {author} {\bibfnamefont {T.~C.}\ \bibnamefont
  {Kobayashi}},\ }\bibfield  {title} {\bibinfo {title} {Phonon-mediated
  superconductivity in the {Sb} square-net compound {LaCuSb$_2$}},\ }\href
  {https://doi.org/10.1103/PhysRevB.107.245117} {\bibfield  {journal} {\bibinfo
   {journal} {Phys. Rev. B}\ }\textbf {\bibinfo {volume} {107}},\ \bibinfo
  {pages} {245117} (\bibinfo {year} {2023})},\ \bibinfo {note} {publisher:
  American Physical Society}\BibitemShut {NoStop}%
\bibitem [{\citenamefont {Du}\ \emph {et~al.}(2020)\citenamefont {Du},
  \citenamefont {Su}, \citenamefont {Luo}, \citenamefont {Shen}, \citenamefont
  {Nie}, \citenamefont {Yin}, \citenamefont {Chen}, \citenamefont {Li},
  \citenamefont {Smidman},\ and\ \citenamefont {Yuan}}]{du_interplay_2020}%
  \BibitemOpen
  \bibfield  {author} {\bibinfo {author} {\bibfnamefont {F.}~\bibnamefont
  {Du}}, \bibinfo {author} {\bibfnamefont {H.}~\bibnamefont {Su}}, \bibinfo
  {author} {\bibfnamefont {S.~S.}\ \bibnamefont {Luo}}, \bibinfo {author}
  {\bibfnamefont {B.}~\bibnamefont {Shen}}, \bibinfo {author} {\bibfnamefont
  {Z.~Y.}\ \bibnamefont {Nie}}, \bibinfo {author} {\bibfnamefont {L.~C.}\
  \bibnamefont {Yin}}, \bibinfo {author} {\bibfnamefont {Y.}~\bibnamefont
  {Chen}}, \bibinfo {author} {\bibfnamefont {R.}~\bibnamefont {Li}}, \bibinfo
  {author} {\bibfnamefont {M.}~\bibnamefont {Smidman}},\ and\ \bibinfo {author}
  {\bibfnamefont {H.~Q.}\ \bibnamefont {Yuan}},\ }\bibfield  {title} {\bibinfo
  {title} {Interplay between charge density wave order and superconductivity in
  {La}{Au}{Sb$_2$} under pressure},\ }\href
  {https://doi.org/10.1103/PhysRevB.102.144510} {\bibfield  {journal} {\bibinfo
   {journal} {Physical Review B}\ }\textbf {\bibinfo {volume} {102}},\ \bibinfo
  {pages} {144510} (\bibinfo {year} {2020})}\BibitemShut {NoStop}%
\bibitem [{\citenamefont {Kuo}\ \emph {et~al.}(2019)\citenamefont {Kuo},
  \citenamefont {Shen}, \citenamefont {Li}, \citenamefont {Quyen},
  \citenamefont {Tzeng}, \citenamefont {Luo}, \citenamefont {Wang},
  \citenamefont {Kuo},\ and\ \citenamefont {Lue}}]{kuo_characterization_2019}%
  \BibitemOpen
  \bibfield  {author} {\bibinfo {author} {\bibfnamefont {C.~N.}\ \bibnamefont
  {Kuo}}, \bibinfo {author} {\bibfnamefont {D.}~\bibnamefont {Shen}}, \bibinfo
  {author} {\bibfnamefont {B.~S.}\ \bibnamefont {Li}}, \bibinfo {author}
  {\bibfnamefont {N.~N.}\ \bibnamefont {Quyen}}, \bibinfo {author}
  {\bibfnamefont {W.~Y.}\ \bibnamefont {Tzeng}}, \bibinfo {author}
  {\bibfnamefont {C.~W.}\ \bibnamefont {Luo}}, \bibinfo {author} {\bibfnamefont
  {L.~M.}\ \bibnamefont {Wang}}, \bibinfo {author} {\bibfnamefont {Y.~K.}\
  \bibnamefont {Kuo}},\ and\ \bibinfo {author} {\bibfnamefont {C.~S.}\
  \bibnamefont {Lue}},\ }\bibfield  {title} {\bibinfo {title} {Characterization
  of the charge density wave transition and observation of the amplitude mode
  in {LaAuSb$_2$}},\ }\href {https://doi.org/10.1103/PhysRevB.99.235121}
  {\bibfield  {journal} {\bibinfo  {journal} {Physical Review B}\ }\textbf
  {\bibinfo {volume} {99}},\ \bibinfo {pages} {235121} (\bibinfo {year}
  {2019})},\ \bibinfo {note} {publisher: American Physical Society}\BibitemShut
  {NoStop}%
\bibitem [{\citenamefont {Xiang}\ \emph {et~al.}(2020)\citenamefont {Xiang},
  \citenamefont {Ryan}, \citenamefont {Straszheim}, \citenamefont {Canfield},\
  and\ \citenamefont {Bud'ko}}]{xiang_tuning_2020}%
  \BibitemOpen
  \bibfield  {author} {\bibinfo {author} {\bibfnamefont {L.}~\bibnamefont
  {Xiang}}, \bibinfo {author} {\bibfnamefont {D.~H.}\ \bibnamefont {Ryan}},
  \bibinfo {author} {\bibfnamefont {W.~E.}\ \bibnamefont {Straszheim}},
  \bibinfo {author} {\bibfnamefont {P.~C.}\ \bibnamefont {Canfield}},\ and\
  \bibinfo {author} {\bibfnamefont {S.~L.}\ \bibnamefont {Bud'ko}},\ }\bibfield
   {title} {\bibinfo {title} {Tuning of charge density wave transitions in
  {LaAu$_x$Sb$_2$} by pressure and {Au} stoichiometry},\ }\href
  {https://doi.org/10.1103/PhysRevB.102.125110} {\bibfield  {journal} {\bibinfo
   {journal} {Physical Review B}\ }\textbf {\bibinfo {volume} {102}},\ \bibinfo
  {pages} {125110} (\bibinfo {year} {2020})}\BibitemShut {NoStop}%
\bibitem [{\citenamefont {Xiang}\ \emph {et~al.}(2022)\citenamefont {Xiang},
  \citenamefont {Ryan}, \citenamefont {Canfield},\ and\ \citenamefont
  {Bud’ko}}]{xiang_effects_2022}%
  \BibitemOpen
  \bibfield  {author} {\bibinfo {author} {\bibfnamefont {L.}~\bibnamefont
  {Xiang}}, \bibinfo {author} {\bibfnamefont {D.~H.}\ \bibnamefont {Ryan}},
  \bibinfo {author} {\bibfnamefont {P.~C.}\ \bibnamefont {Canfield}},\ and\
  \bibinfo {author} {\bibfnamefont {S.~L.}\ \bibnamefont {Bud’ko}},\
  }\bibfield  {title} {\bibinfo {title} {Effects of {Physical} and {Chemical}
  {Pressure} on {Charge} {Density} {Wave} {Transitions} in
  {LaAg$_{1-x}$Au$_x$Sb$_2$} {Single} {Crystals}},\ }\bibfield  {journal}
  {\bibinfo  {journal} {Crystals}\ }\textbf {\bibinfo {volume} {12}},\ \href
  {https://doi.org/10.3390/cryst12121693} {10.3390/cryst12121693} (\bibinfo
  {year} {2022})\BibitemShut {NoStop}%
\bibitem [{\citenamefont {Myers}\ \emph {et~al.}(1999)\citenamefont {Myers},
  \citenamefont {Bud’ko}, \citenamefont {Antropov}, \citenamefont {Harmon},
  \citenamefont {Canfield},\ and\ \citenamefont
  {Lacerda}}]{myers_haasvan_1999}%
  \BibitemOpen
  \bibfield  {author} {\bibinfo {author} {\bibfnamefont {K.~D.}\ \bibnamefont
  {Myers}}, \bibinfo {author} {\bibfnamefont {S.~L.}\ \bibnamefont {Bud’ko}},
  \bibinfo {author} {\bibfnamefont {V.~P.}\ \bibnamefont {Antropov}}, \bibinfo
  {author} {\bibfnamefont {B.~N.}\ \bibnamefont {Harmon}}, \bibinfo {author}
  {\bibfnamefont {P.~C.}\ \bibnamefont {Canfield}},\ and\ \bibinfo {author}
  {\bibfnamefont {A.~H.}\ \bibnamefont {Lacerda}},\ }\bibfield  {title}
  {\bibinfo {title} {de {Haas}–van {Alphen} and {Shubnikov}–de {Haas}
  oscillations in {$R$AgSb$_2$} ( {$R$} = {Y} , {La}-{Nd}, {Sm})},\ }\href
  {https://doi.org/10.1103/PhysRevB.60.13371} {\bibfield  {journal} {\bibinfo
  {journal} {Physical Review B}\ }\textbf {\bibinfo {volume} {60}},\ \bibinfo
  {pages} {13371} (\bibinfo {year} {1999})}\BibitemShut {NoStop}%
\bibitem [{\citenamefont {Kresse}\ and\ \citenamefont
  {Furthmüller}(1996{\natexlab{a}})}]{kresse_efficient_1996}%
  \BibitemOpen
  \bibfield  {author} {\bibinfo {author} {\bibfnamefont {G.}~\bibnamefont
  {Kresse}}\ and\ \bibinfo {author} {\bibfnamefont {J.}~\bibnamefont
  {Furthmüller}},\ }\bibfield  {title} {\bibinfo {title} {Efficient iterative
  schemes for {$ab$} {$initio$} total-energy calculations using a plane-wave
  basis set},\ }\href {https://doi.org/10.1103/PhysRevB.54.11169} {\bibfield
  {journal} {\bibinfo  {journal} {Physical Review B}\ }\textbf {\bibinfo
  {volume} {54}},\ \bibinfo {pages} {11169} (\bibinfo {year}
  {1996}{\natexlab{a}})},\ \bibinfo {note} {publisher: American Physical
  Society}\BibitemShut {NoStop}%
\bibitem [{\citenamefont {Kresse}\ and\ \citenamefont
  {Furthmüller}(1996{\natexlab{b}})}]{kresse_efficiency_1996}%
  \BibitemOpen
  \bibfield  {author} {\bibinfo {author} {\bibfnamefont {G.}~\bibnamefont
  {Kresse}}\ and\ \bibinfo {author} {\bibfnamefont {J.}~\bibnamefont
  {Furthmüller}},\ }\bibfield  {title} {\bibinfo {title} {Efficiency of
  {$ab-initio$} total energy calculations for metals and semiconductors using a
  plane-wave basis set},\ }\href {https://doi.org/10.1016/0927-0256(96)00008-0}
  {\bibfield  {journal} {\bibinfo  {journal} {Computational Materials Science}\
  }\textbf {\bibinfo {volume} {6}},\ \bibinfo {pages} {15} (\bibinfo {year}
  {1996}{\natexlab{b}})}\BibitemShut {NoStop}%
\bibitem [{\citenamefont {Perdew}\ \emph {et~al.}(1996)\citenamefont {Perdew},
  \citenamefont {Burke},\ and\ \citenamefont
  {Ernzerhof}}]{perdew_generalized_1996}%
  \BibitemOpen
  \bibfield  {author} {\bibinfo {author} {\bibfnamefont {J.~P.}\ \bibnamefont
  {Perdew}}, \bibinfo {author} {\bibfnamefont {K.}~\bibnamefont {Burke}},\ and\
  \bibinfo {author} {\bibfnamefont {M.}~\bibnamefont {Ernzerhof}},\ }\bibfield
  {title} {\bibinfo {title} {Generalized {Gradient} {Approximation} {Made}
  {Simple}},\ }\href {https://doi.org/10.1103/PhysRevLett.77.3865} {\bibfield
  {journal} {\bibinfo  {journal} {Physical Review Letters}\ }\textbf {\bibinfo
  {volume} {77}},\ \bibinfo {pages} {3865} (\bibinfo {year} {1996})},\ \bibinfo
  {note} {publisher: American Physical Society}\BibitemShut {NoStop}%
\bibitem [{\citenamefont {Blöchl}(1994)}]{blochl_projector_1994}%
  \BibitemOpen
  \bibfield  {author} {\bibinfo {author} {\bibfnamefont {P.~E.}\ \bibnamefont
  {Blöchl}},\ }\bibfield  {title} {\bibinfo {title} {Projector augmented-wave
  method},\ }\href {https://doi.org/10.1103/PhysRevB.50.17953} {\bibfield
  {journal} {\bibinfo  {journal} {Physical Review B}\ }\textbf {\bibinfo
  {volume} {50}},\ \bibinfo {pages} {17953} (\bibinfo {year} {1994})},\
  \bibinfo {note} {publisher: American Physical Society}\BibitemShut {NoStop}%
\bibitem [{\citenamefont {Ganose}\ \emph {et~al.}(2021)\citenamefont {Ganose},
  \citenamefont {Searle}, \citenamefont {Jain},\ and\ \citenamefont
  {Griffin}}]{ganose_ifermi_2021}%
  \BibitemOpen
  \bibfield  {author} {\bibinfo {author} {\bibfnamefont {A.}~\bibnamefont
  {Ganose}}, \bibinfo {author} {\bibfnamefont {A.}~\bibnamefont {Searle}},
  \bibinfo {author} {\bibfnamefont {A.}~\bibnamefont {Jain}},\ and\ \bibinfo
  {author} {\bibfnamefont {S.}~\bibnamefont {Griffin}},\ }\bibfield  {title}
  {\bibinfo {title} {{IFermi}: {A} python library for {Fermi} surface
  generation and analysis},\ }\href {https://doi.org/10.21105/joss.03089}
  {\bibfield  {journal} {\bibinfo  {journal} {Journal of Open Source Software}\
  }\textbf {\bibinfo {volume} {6}},\ \bibinfo {pages} {3089} (\bibinfo {year}
  {2021})}\BibitemShut {NoStop}%
\bibitem [{\citenamefont {Prakash}\ \emph {et~al.}(2016)\citenamefont
  {Prakash}, \citenamefont {Thamizhavel},\ and\ \citenamefont
  {Ramakrishnan}}]{prakash_ferromagnetic_2016}%
  \BibitemOpen
  \bibfield  {author} {\bibinfo {author} {\bibfnamefont {O.}~\bibnamefont
  {Prakash}}, \bibinfo {author} {\bibfnamefont {A.}~\bibnamefont
  {Thamizhavel}},\ and\ \bibinfo {author} {\bibfnamefont {S.}~\bibnamefont
  {Ramakrishnan}},\ }\bibfield  {title} {\bibinfo {title} {Ferromagnetic
  ordering of minority {Ce$^{3+}$} spins in a quasi-skutterudite
  {Ce$_3$Os$_4$Ge$_{13}$} single crystal},\ }\href
  {https://doi.org/10.1103/PhysRevB.93.064427} {\bibfield  {journal} {\bibinfo
  {journal} {Physical Review B}\ }\textbf {\bibinfo {volume} {93}},\ \bibinfo
  {pages} {064427} (\bibinfo {year} {2016})}\BibitemShut {NoStop}%
\bibitem [{\citenamefont {Teng}\ \emph {et~al.}(2022)\citenamefont {Teng},
  \citenamefont {Chen}, \citenamefont {Ye}, \citenamefont {Rosenberg},
  \citenamefont {Liu}, \citenamefont {Yin}, \citenamefont {Jiang},
  \citenamefont {Oh}, \citenamefont {Hasan}, \citenamefont {Neubauer},
  \citenamefont {Gao}, \citenamefont {Xie}, \citenamefont {Hashimoto},
  \citenamefont {Lu}, \citenamefont {Jozwiak}, \citenamefont {Bostwick},
  \citenamefont {Rotenberg}, \citenamefont {Birgeneau}, \citenamefont {Chu},
  \citenamefont {Yi},\ and\ \citenamefont {Dai}}]{teng_discovery_2022}%
  \BibitemOpen
  \bibfield  {author} {\bibinfo {author} {\bibfnamefont {X.}~\bibnamefont
  {Teng}}, \bibinfo {author} {\bibfnamefont {L.}~\bibnamefont {Chen}}, \bibinfo
  {author} {\bibfnamefont {F.}~\bibnamefont {Ye}}, \bibinfo {author}
  {\bibfnamefont {E.}~\bibnamefont {Rosenberg}}, \bibinfo {author}
  {\bibfnamefont {Z.}~\bibnamefont {Liu}}, \bibinfo {author} {\bibfnamefont
  {J.-X.}\ \bibnamefont {Yin}}, \bibinfo {author} {\bibfnamefont {Y.-X.}\
  \bibnamefont {Jiang}}, \bibinfo {author} {\bibfnamefont {J.~S.}\ \bibnamefont
  {Oh}}, \bibinfo {author} {\bibfnamefont {M.~Z.}\ \bibnamefont {Hasan}},
  \bibinfo {author} {\bibfnamefont {K.~J.}\ \bibnamefont {Neubauer}}, \bibinfo
  {author} {\bibfnamefont {B.}~\bibnamefont {Gao}}, \bibinfo {author}
  {\bibfnamefont {Y.}~\bibnamefont {Xie}}, \bibinfo {author} {\bibfnamefont
  {M.}~\bibnamefont {Hashimoto}}, \bibinfo {author} {\bibfnamefont
  {D.}~\bibnamefont {Lu}}, \bibinfo {author} {\bibfnamefont {C.}~\bibnamefont
  {Jozwiak}}, \bibinfo {author} {\bibfnamefont {A.}~\bibnamefont {Bostwick}},
  \bibinfo {author} {\bibfnamefont {E.}~\bibnamefont {Rotenberg}}, \bibinfo
  {author} {\bibfnamefont {R.~J.}\ \bibnamefont {Birgeneau}}, \bibinfo {author}
  {\bibfnamefont {J.-H.}\ \bibnamefont {Chu}}, \bibinfo {author} {\bibfnamefont
  {M.}~\bibnamefont {Yi}},\ and\ \bibinfo {author} {\bibfnamefont
  {P.}~\bibnamefont {Dai}},\ }\bibfield  {title} {\bibinfo {title} {Discovery
  of charge density wave in a kagome lattice antiferromagnet},\ }\href
  {https://doi.org/10.1038/s41586-022-05034-z} {\bibfield  {journal} {\bibinfo
  {journal} {Nature}\ }\textbf {\bibinfo {volume} {609}},\ \bibinfo {pages}
  {490} (\bibinfo {year} {2022})}\BibitemShut {NoStop}%
\bibitem [{\citenamefont {Lue}\ \emph {et~al.}(2007)\citenamefont {Lue},
  \citenamefont {Tao}, \citenamefont {Sivakumar},\ and\ \citenamefont
  {Kuo}}]{lue_weak_2007}%
  \BibitemOpen
  \bibfield  {author} {\bibinfo {author} {\bibfnamefont {C.~S.}\ \bibnamefont
  {Lue}}, \bibinfo {author} {\bibfnamefont {Y.~F.}\ \bibnamefont {Tao}},
  \bibinfo {author} {\bibfnamefont {K.~M.}\ \bibnamefont {Sivakumar}},\ and\
  \bibinfo {author} {\bibfnamefont {Y.~K.}\ \bibnamefont {Kuo}},\ }\bibfield
  {title} {\bibinfo {title} {Weak charge-density-wave transition in
  {LaAgSb$_2$} investigated by transport, thermal, and {NMR} studies},\ }\href
  {https://doi.org/10.1088/0953-8984/19/40/406230} {\bibfield  {journal}
  {\bibinfo  {journal} {Journal of Physics: Condensed Matter}\ }\textbf
  {\bibinfo {volume} {19}},\ \bibinfo {pages} {406230} (\bibinfo {year}
  {2007})}\BibitemShut {NoStop}%
\bibitem [{\citenamefont {Pippard}(1989)}]{pippard_magnetoresistance_1989}%
  \BibitemOpen
  \bibfield  {author} {\bibinfo {author} {\bibfnamefont {A.}~\bibnamefont
  {Pippard}},\ }\href {https://books.google.com.hk/books?id=D5XHMARd2ocC}
  {\emph {\bibinfo {title} {Magnetoresistance in {Metals}}}},\ Cambridge
  {Studies} in {Low} {Temperature} {Physics}\ (\bibinfo  {publisher} {Cambridge
  University Press},\ \bibinfo {year} {1989})\BibitemShut {NoStop}%
\bibitem [{\citenamefont {Pletikosi{\'{c}}}\ \emph {et~al.}(2014)\citenamefont
  {Pletikosi{\'{c}}}, \citenamefont {Ali}, \citenamefont {Fedorov},
  \citenamefont {Cava},\ and\ \citenamefont {Valla}}]{Pletikosic2014}%
  \BibitemOpen
  \bibfield  {author} {\bibinfo {author} {\bibfnamefont {I.}~\bibnamefont
  {Pletikosi{\'{c}}}}, \bibinfo {author} {\bibfnamefont {M.~N.}\ \bibnamefont
  {Ali}}, \bibinfo {author} {\bibfnamefont {A.~V.}\ \bibnamefont {Fedorov}},
  \bibinfo {author} {\bibfnamefont {R.~J.}\ \bibnamefont {Cava}},\ and\
  \bibinfo {author} {\bibfnamefont {T.}~\bibnamefont {Valla}},\ }\bibfield
  {title} {\bibinfo {title} {{Electronic structure basis for the extraordinary
  magnetoresistance in {WTe$_2$}}},\ }\href
  {https://doi.org/10.1103/PhysRevLett.113.216601} {\bibfield  {journal}
  {\bibinfo  {journal} {Physical Review Letters}\ }\textbf {\bibinfo {volume}
  {113}},\ \bibinfo {pages} {216601} (\bibinfo {year} {2014})}\BibitemShut
  {NoStop}%
\bibitem [{\citenamefont {Stroud}\ and\ \citenamefont
  {Pan}(1979)}]{stroud_magnetoresistance_1979}%
  \BibitemOpen
  \bibfield  {author} {\bibinfo {author} {\bibfnamefont {D.}~\bibnamefont
  {Stroud}}\ and\ \bibinfo {author} {\bibfnamefont {F.~P.}\ \bibnamefont
  {Pan}},\ }\bibfield  {title} {\bibinfo {title} {Magnetoresistance and {Hall}
  coefficient of inhomogeneous metals},\ }\href
  {https://doi.org/10.1103/PhysRevB.20.455} {\bibfield  {journal} {\bibinfo
  {journal} {Physical Review B}\ }\textbf {\bibinfo {volume} {20}},\ \bibinfo
  {pages} {455} (\bibinfo {year} {1979})}\BibitemShut {NoStop}%
\bibitem [{\citenamefont {Hu}\ and\ \citenamefont
  {Rosenbaum}(2008)}]{hu_classical_2008}%
  \BibitemOpen
  \bibfield  {author} {\bibinfo {author} {\bibfnamefont {J.}~\bibnamefont
  {Hu}}\ and\ \bibinfo {author} {\bibfnamefont {T.~F.}\ \bibnamefont
  {Rosenbaum}},\ }\bibfield  {title} {\bibinfo {title} {Classical and quantum
  routes to linear magnetoresistance},\ }\href
  {https://doi.org/10.1038/nmat2259} {\bibfield  {journal} {\bibinfo  {journal}
  {Nature Materials}\ }\textbf {\bibinfo {volume} {7}},\ \bibinfo {pages} {697}
  (\bibinfo {year} {2008})},\ \bibinfo {note} {number: 9 Publisher: Nature
  Publishing Group}\BibitemShut {NoStop}%
\bibitem [{\citenamefont {Kohler}(1938)}]{kohler_zur_1938}%
  \BibitemOpen
  \bibfield  {author} {\bibinfo {author} {\bibfnamefont {M.}~\bibnamefont
  {Kohler}},\ }\bibfield  {title} {\bibinfo {title} {Zur magnetischen
  {Widerstandsänderung} reiner {Metalle}},\ }\href
  {https://doi.org/10.1002/andp.19384240124} {\bibfield  {journal} {\bibinfo
  {journal} {Annalen der Physik}\ }\textbf {\bibinfo {volume} {424}},\ \bibinfo
  {pages} {211} (\bibinfo {year} {1938})}\BibitemShut {NoStop}%
\bibitem [{\citenamefont {McKenzie}\ \emph {et~al.}(1998)\citenamefont
  {McKenzie}, \citenamefont {Qualls}, \citenamefont {Han},\ and\ \citenamefont
  {Brooks}}]{mckenzie_violation_1998}%
  \BibitemOpen
  \bibfield  {author} {\bibinfo {author} {\bibfnamefont {R.~H.}\ \bibnamefont
  {McKenzie}}, \bibinfo {author} {\bibfnamefont {J.~S.}\ \bibnamefont
  {Qualls}}, \bibinfo {author} {\bibfnamefont {S.~Y.}\ \bibnamefont {Han}},\
  and\ \bibinfo {author} {\bibfnamefont {J.~S.}\ \bibnamefont {Brooks}},\
  }\bibfield  {title} {\bibinfo {title} {Violation of {Kohler}'s rule by the
  magnetoresistance of a quasi-two-dimensional organic metal},\ }\href
  {https://doi.org/10.1103/PhysRevB.57.11854} {\bibfield  {journal} {\bibinfo
  {journal} {Physical Review B}\ }\textbf {\bibinfo {volume} {57}},\ \bibinfo
  {pages} {11854} (\bibinfo {year} {1998})},\ \bibinfo {note} {publisher:
  American Physical Society}\BibitemShut {NoStop}%
\bibitem [{\citenamefont {Knowles}\ \emph {et~al.}(2020)\citenamefont
  {Knowles}, \citenamefont {Yang}, \citenamefont {Muramatsu}, \citenamefont
  {Moulding}, \citenamefont {Buhot}, \citenamefont {Sayers}, \citenamefont
  {Da~Como},\ and\ \citenamefont {Friedemann}}]{knowles_fermi_2020}%
  \BibitemOpen
  \bibfield  {author} {\bibinfo {author} {\bibfnamefont {P.}~\bibnamefont
  {Knowles}}, \bibinfo {author} {\bibfnamefont {B.}~\bibnamefont {Yang}},
  \bibinfo {author} {\bibfnamefont {T.}~\bibnamefont {Muramatsu}}, \bibinfo
  {author} {\bibfnamefont {O.}~\bibnamefont {Moulding}}, \bibinfo {author}
  {\bibfnamefont {J.}~\bibnamefont {Buhot}}, \bibinfo {author} {\bibfnamefont
  {C.~J.}\ \bibnamefont {Sayers}}, \bibinfo {author} {\bibfnamefont
  {E.}~\bibnamefont {Da~Como}},\ and\ \bibinfo {author} {\bibfnamefont
  {S.}~\bibnamefont {Friedemann}},\ }\bibfield  {title} {\bibinfo {title}
  {Fermi {Surface} {Reconstruction} and {Electron} {Dynamics} at the
  {Charge}-{Density}-{Wave} {Transition} in {TiSe$_2$}},\ }\href
  {https://doi.org/10.1103/PhysRevLett.124.167602} {\bibfield  {journal}
  {\bibinfo  {journal} {Physical Review Letters}\ }\textbf {\bibinfo {volume}
  {124}},\ \bibinfo {pages} {167602} (\bibinfo {year} {2020})}\BibitemShut
  {NoStop}%
\bibitem [{\citenamefont {Noto}\ \emph {et~al.}(1980)\citenamefont {Noto},
  \citenamefont {Morohashi}, \citenamefont {Arikawa},\ and\ \citenamefont
  {Muto}}]{noto_temperature_1980}%
  \BibitemOpen
  \bibfield  {author} {\bibinfo {author} {\bibfnamefont {K.}~\bibnamefont
  {Noto}}, \bibinfo {author} {\bibfnamefont {S.}~\bibnamefont {Morohashi}},
  \bibinfo {author} {\bibfnamefont {K.}~\bibnamefont {Arikawa}},\ and\ \bibinfo
  {author} {\bibfnamefont {Y.}~\bibnamefont {Muto}},\ }\bibfield  {title}
  {\bibinfo {title} {Temperature and magnetic field dependence of the
  electrical resistance in pure and {Fe}-doped {2H}-{NbSe$_2$}},\ }\href
  {https://doi.org/10.1016/0378-4363(80)90233-8} {\bibfield  {journal}
  {\bibinfo  {journal} {Physica B+C}\ }\textbf {\bibinfo {volume} {99}},\
  \bibinfo {pages} {204} (\bibinfo {year} {1980})}\BibitemShut {NoStop}%
\bibitem [{\citenamefont {Smith}(1978)}]{smith_semiconductors_1978}%
  \BibitemOpen
  \bibfield  {author} {\bibinfo {author} {\bibfnamefont {R.~A.}\ \bibnamefont
  {Smith}},\ }\href@noop {} {\emph {\bibinfo {title} {Semiconductors}}},\
  \bibinfo {edition} {2nd}\ ed.\ (\bibinfo  {publisher} {New York: Cambridge
  University Press},\ \bibinfo {year} {1978})\BibitemShut {NoStop}%
\bibitem [{\citenamefont {Shoenberg}(1984)}]{shoenberg_1984}%
  \BibitemOpen
  \bibfield  {author} {\bibinfo {author} {\bibfnamefont {D.}~\bibnamefont
  {Shoenberg}},\ }\href {https://doi.org/10.1017/CBO9780511897870} {\emph
  {\bibinfo {title} {Magnetic Oscillations in Metals}}},\ Cambridge Monographs
  on Physics\ (\bibinfo  {publisher} {Cambridge University Press},\ \bibinfo
  {year} {1984})\BibitemShut {NoStop}%
\bibitem [{\citenamefont {He}\ \emph {et~al.}(2017)\citenamefont {He},
  \citenamefont {Fu}, \citenamefont {Zhao}, \citenamefont {Liang},
  \citenamefont {Chen}, \citenamefont {Leng}, \citenamefont {Wang},
  \citenamefont {Li}, \citenamefont {Zhang}, \citenamefont {Xue}, \citenamefont
  {Li}, \citenamefont {Zhang}, \citenamefont {Ren},\ and\ \citenamefont
  {Chen}}]{he_quasi-two-dimensional_2017}%
  \BibitemOpen
  \bibfield  {author} {\bibinfo {author} {\bibfnamefont {J.~B.}\ \bibnamefont
  {He}}, \bibinfo {author} {\bibfnamefont {Y.}~\bibnamefont {Fu}}, \bibinfo
  {author} {\bibfnamefont {L.~X.}\ \bibnamefont {Zhao}}, \bibinfo {author}
  {\bibfnamefont {H.}~\bibnamefont {Liang}}, \bibinfo {author} {\bibfnamefont
  {D.}~\bibnamefont {Chen}}, \bibinfo {author} {\bibfnamefont {Y.~M.}\
  \bibnamefont {Leng}}, \bibinfo {author} {\bibfnamefont {X.~M.}\ \bibnamefont
  {Wang}}, \bibinfo {author} {\bibfnamefont {J.}~\bibnamefont {Li}}, \bibinfo
  {author} {\bibfnamefont {S.}~\bibnamefont {Zhang}}, \bibinfo {author}
  {\bibfnamefont {M.~Q.}\ \bibnamefont {Xue}}, \bibinfo {author} {\bibfnamefont
  {C.~H.}\ \bibnamefont {Li}}, \bibinfo {author} {\bibfnamefont
  {P.}~\bibnamefont {Zhang}}, \bibinfo {author} {\bibfnamefont {Z.~A.}\
  \bibnamefont {Ren}},\ and\ \bibinfo {author} {\bibfnamefont {G.~F.}\
  \bibnamefont {Chen}},\ }\bibfield  {title} {\bibinfo {title}
  {Quasi-two-dimensional massless {Dirac} fermions in {CaMnSb$_2$}},\ }\href
  {https://doi.org/10.1103/PhysRevB.95.045128} {\bibfield  {journal} {\bibinfo
  {journal} {Physical Review B}\ }\textbf {\bibinfo {volume} {95}},\ \bibinfo
  {pages} {045128} (\bibinfo {year} {2017})}\BibitemShut {NoStop}%
\bibitem [{\citenamefont {Liu}\ \emph {et~al.}(2017)\citenamefont {Liu},
  \citenamefont {Hu}, \citenamefont {Zhang}, \citenamefont {Graf},
  \citenamefont {Cao}, \citenamefont {Radmanesh}, \citenamefont {Adams},
  \citenamefont {Zhu}, \citenamefont {Cheng}, \citenamefont {Liu},
  \citenamefont {Phelan}, \citenamefont {Wei}, \citenamefont {Jaime},
  \citenamefont {Balakirev}, \citenamefont {Tennant}, \citenamefont {DiTusa},
  \citenamefont {Chiorescu}, \citenamefont {Spinu},\ and\ \citenamefont
  {Mao}}]{liu_magnetic_2017}%
  \BibitemOpen
  \bibfield  {author} {\bibinfo {author} {\bibfnamefont {J.~Y.}\ \bibnamefont
  {Liu}}, \bibinfo {author} {\bibfnamefont {J.}~\bibnamefont {Hu}}, \bibinfo
  {author} {\bibfnamefont {Q.}~\bibnamefont {Zhang}}, \bibinfo {author}
  {\bibfnamefont {D.}~\bibnamefont {Graf}}, \bibinfo {author} {\bibfnamefont
  {H.~B.}\ \bibnamefont {Cao}}, \bibinfo {author} {\bibfnamefont {S.~M.~A.}\
  \bibnamefont {Radmanesh}}, \bibinfo {author} {\bibfnamefont {D.~J.}\
  \bibnamefont {Adams}}, \bibinfo {author} {\bibfnamefont {Y.~L.}\ \bibnamefont
  {Zhu}}, \bibinfo {author} {\bibfnamefont {G.}~\bibnamefont {Cheng}}, \bibinfo
  {author} {\bibfnamefont {X.}~\bibnamefont {Liu}}, \bibinfo {author}
  {\bibfnamefont {W.~A.}\ \bibnamefont {Phelan}}, \bibinfo {author}
  {\bibfnamefont {J.}~\bibnamefont {Wei}}, \bibinfo {author} {\bibfnamefont
  {M.}~\bibnamefont {Jaime}}, \bibinfo {author} {\bibfnamefont
  {F.}~\bibnamefont {Balakirev}}, \bibinfo {author} {\bibfnamefont {D.~A.}\
  \bibnamefont {Tennant}}, \bibinfo {author} {\bibfnamefont {J.~F.}\
  \bibnamefont {DiTusa}}, \bibinfo {author} {\bibfnamefont {I.}~\bibnamefont
  {Chiorescu}}, \bibinfo {author} {\bibfnamefont {L.}~\bibnamefont {Spinu}},\
  and\ \bibinfo {author} {\bibfnamefont {Z.~Q.}\ \bibnamefont {Mao}},\
  }\bibfield  {title} {\bibinfo {title} {A magnetic topological semimetal
  {Sr$_{1-y}$Mn$_{1-z}$Sb$_2$} ($y$, $z$ {\textless} 0.1)},\ }\href
  {https://doi.org/10.1038/nmat4953} {\bibfield  {journal} {\bibinfo  {journal}
  {Nature Materials}\ }\textbf {\bibinfo {volume} {16}},\ \bibinfo {pages}
  {905} (\bibinfo {year} {2017})}\BibitemShut {NoStop}%
\bibitem [{\citenamefont {Huang}\ \emph {et~al.}(2017)\citenamefont {Huang},
  \citenamefont {Kim}, \citenamefont {Shelton}, \citenamefont {Plummer},\ and\
  \citenamefont {Jin}}]{huang_nontrivial_2017}%
  \BibitemOpen
  \bibfield  {author} {\bibinfo {author} {\bibfnamefont {S.}~\bibnamefont
  {Huang}}, \bibinfo {author} {\bibfnamefont {J.}~\bibnamefont {Kim}}, \bibinfo
  {author} {\bibfnamefont {W.~A.}\ \bibnamefont {Shelton}}, \bibinfo {author}
  {\bibfnamefont {E.~W.}\ \bibnamefont {Plummer}},\ and\ \bibinfo {author}
  {\bibfnamefont {R.}~\bibnamefont {Jin}},\ }\bibfield  {title} {\bibinfo
  {title} {Nontrivial {Berry} phase in magnetic {BaMnSb$_2$} semimetal},\
  }\href {https://doi.org/10.1073/pnas.1706657114} {\bibfield  {journal}
  {\bibinfo  {journal} {Proceedings of the National Academy of Sciences}\
  }\textbf {\bibinfo {volume} {114}},\ \bibinfo {pages} {6256} (\bibinfo {year}
  {2017})}\BibitemShut {NoStop}%
\bibitem [{\citenamefont {Wang}\ \emph {et~al.}(2018)\citenamefont {Wang},
  \citenamefont {Xu}, \citenamefont {Sun},\ and\ \citenamefont
  {Xia}}]{wang_quantum_2018}%
  \BibitemOpen
  \bibfield  {author} {\bibinfo {author} {\bibfnamefont {Y.-Y.}\ \bibnamefont
  {Wang}}, \bibinfo {author} {\bibfnamefont {S.}~\bibnamefont {Xu}}, \bibinfo
  {author} {\bibfnamefont {L.-L.}\ \bibnamefont {Sun}},\ and\ \bibinfo {author}
  {\bibfnamefont {T.-L.}\ \bibnamefont {Xia}},\ }\bibfield  {title} {\bibinfo
  {title} {Quantum oscillations and coherent interlayer transport in a new
  topological {Dirac} semimetal candidate {YbMnSb$_2$}},\ }\href
  {https://doi.org/10.1103/PhysRevMaterials.2.021201} {\bibfield  {journal}
  {\bibinfo  {journal} {Physical Review Materials}\ }\textbf {\bibinfo {volume}
  {2}},\ \bibinfo {pages} {021201} (\bibinfo {year} {2018})}\BibitemShut
  {NoStop}%
\bibitem [{\citenamefont {Hu}\ \emph {et~al.}(2019)\citenamefont {Hu},
  \citenamefont {Xu}, \citenamefont {Ni},\ and\ \citenamefont
  {Mao}}]{hu_transport_2019}%
  \BibitemOpen
  \bibfield  {author} {\bibinfo {author} {\bibfnamefont {J.}~\bibnamefont
  {Hu}}, \bibinfo {author} {\bibfnamefont {S.-Y.}\ \bibnamefont {Xu}}, \bibinfo
  {author} {\bibfnamefont {N.}~\bibnamefont {Ni}},\ and\ \bibinfo {author}
  {\bibfnamefont {Z.}~\bibnamefont {Mao}},\ }\bibfield  {title} {\bibinfo
  {title} {Transport of topological semimetals},\ }\href
  {https://doi.org/10.1146/annurev-matsci-070218-010023} {\bibfield  {journal}
  {\bibinfo  {journal} {Annual Review of Materials Research}\ }\textbf
  {\bibinfo {volume} {49}},\ \bibinfo {pages} {207} (\bibinfo {year}
  {2019})}\BibitemShut {NoStop}%
\bibitem [{SM(2023)}]{SM}%
  \BibitemOpen
  \href@noop {} {}\bibinfo {howpublished} {See Supplemental Material at [URL
  will be inserted by publisher] for Magnetization, DOS, and more details of
  calculated Fermi surface.} (\bibinfo {year} {2023})\BibitemShut {NoStop}%
\bibitem [{\citenamefont {Rosmus}\ \emph {et~al.}(2022)\citenamefont {Rosmus},
  \citenamefont {Olszowska}, \citenamefont {Bukowski}, \citenamefont
  {Starowicz}, \citenamefont {Piekarz},\ and\ \citenamefont
  {Ptok}}]{rosmus_electronic_2022}%
  \BibitemOpen
  \bibfield  {author} {\bibinfo {author} {\bibfnamefont {M.}~\bibnamefont
  {Rosmus}}, \bibinfo {author} {\bibfnamefont {N.}~\bibnamefont {Olszowska}},
  \bibinfo {author} {\bibfnamefont {Z.}~\bibnamefont {Bukowski}}, \bibinfo
  {author} {\bibfnamefont {P.}~\bibnamefont {Starowicz}}, \bibinfo {author}
  {\bibfnamefont {P.}~\bibnamefont {Piekarz}},\ and\ \bibinfo {author}
  {\bibfnamefont {A.}~\bibnamefont {Ptok}},\ }\bibfield  {title} {\bibinfo
  {title} {Electronic {Band} {Structure} and {Surface} {States} in {Dirac}
  {Semimetal} {LaAgSb$_2$}},\ }\href {https://doi.org/10.3390/ma15207168}
  {\bibfield  {journal} {\bibinfo  {journal} {Materials}\ }\textbf {\bibinfo
  {volume} {15}},\ \bibinfo {pages} {7168} (\bibinfo {year} {2022})},\ \bibinfo
  {note} {number: 20 Publisher: Multidisciplinary Digital Publishing
  Institute}\BibitemShut {NoStop}%
\bibitem [{\citenamefont {Hase}\ and\ \citenamefont
  {Yanagisawa}(2014)}]{hase_electronic_2014}%
  \BibitemOpen
  \bibfield  {author} {\bibinfo {author} {\bibfnamefont {I.}~\bibnamefont
  {Hase}}\ and\ \bibinfo {author} {\bibfnamefont {T.}~\bibnamefont
  {Yanagisawa}},\ }\bibfield  {title} {\bibinfo {title} {Electronic {Band}
  {Calculation} of {La$T$Sb$_2$} ({$T$} = {Cu},{Ag},{Au})},\ }\href
  {https://doi.org/10.1016/j.phpro.2014.09.011} {\bibfield  {journal} {\bibinfo
   {journal} {Physics Procedia}\ }\bibinfo {series} {Proceedings of the 26th
  {International} {Symposium} on {Superconductivity} ({ISS} 2013)},\ \textbf
  {\bibinfo {volume} {58}},\ \bibinfo {pages} {42} (\bibinfo {year}
  {2014})}\BibitemShut {NoStop}%
\bibitem [{\citenamefont {Valla}\ \emph {et~al.}(2004)\citenamefont {Valla},
  \citenamefont {Fedorov}, \citenamefont {Johnson}, \citenamefont {Glans},
  \citenamefont {McGuinness}, \citenamefont {Smith}, \citenamefont {Andrei},\
  and\ \citenamefont {Berger}}]{2H_NbSe2}%
  \BibitemOpen
  \bibfield  {author} {\bibinfo {author} {\bibfnamefont {T.}~\bibnamefont
  {Valla}}, \bibinfo {author} {\bibfnamefont {A.~V.}\ \bibnamefont {Fedorov}},
  \bibinfo {author} {\bibfnamefont {P.~D.}\ \bibnamefont {Johnson}}, \bibinfo
  {author} {\bibfnamefont {P.-A.}\ \bibnamefont {Glans}}, \bibinfo {author}
  {\bibfnamefont {C.}~\bibnamefont {McGuinness}}, \bibinfo {author}
  {\bibfnamefont {K.~E.}\ \bibnamefont {Smith}}, \bibinfo {author}
  {\bibfnamefont {E.~Y.}\ \bibnamefont {Andrei}},\ and\ \bibinfo {author}
  {\bibfnamefont {H.}~\bibnamefont {Berger}},\ }\bibfield  {title} {\bibinfo
  {title} {Quasiparticle {Spectra}, {Charge}-{Density} {Waves},
  {Superconductivity}, and {Electron}-{Phonon} {Coupling} in {2H-NbSe$_2$}},\
  }\href {https://doi.org/10.1103/PhysRevLett.92.086401} {\bibfield  {journal}
  {\bibinfo  {journal} {Physical Review Letters}\ }\textbf {\bibinfo {volume}
  {92}},\ \bibinfo {pages} {086401} (\bibinfo {year} {2004})}\BibitemShut
  {NoStop}%
\bibitem [{\citenamefont {Valla}\ \emph {et~al.}(2006)\citenamefont {Valla},
  \citenamefont {Fedorov}, \citenamefont {Lee}, \citenamefont {Davis},\ and\
  \citenamefont {Gu}}]{Valla2006}%
  \BibitemOpen
  \bibfield  {author} {\bibinfo {author} {\bibfnamefont {T.}~\bibnamefont
  {Valla}}, \bibinfo {author} {\bibfnamefont {A.~V.}\ \bibnamefont {Fedorov}},
  \bibinfo {author} {\bibfnamefont {J.}~\bibnamefont {Lee}}, \bibinfo {author}
  {\bibfnamefont {J.~C.}\ \bibnamefont {Davis}},\ and\ \bibinfo {author}
  {\bibfnamefont {G.~D.}\ \bibnamefont {Gu}},\ }\bibfield  {title} {\bibinfo
  {title} {{The ground state of the pseudogap in cuprate superconductors.}},\
  }\href {https://doi.org/10.1126/science.1134742} {\bibfield  {journal}
  {\bibinfo  {journal} {Science (New York, N.Y.)}\ }\textbf {\bibinfo {volume}
  {314}},\ \bibinfo {pages} {1914} (\bibinfo {year} {2006})}\BibitemShut
  {NoStop}%
\bibitem [{\citenamefont {Feng}\ \emph {et~al.}(2015)\citenamefont {Feng},
  \citenamefont {Wang}, \citenamefont {Chen}, \citenamefont {Shi},
  \citenamefont {Xie}, \citenamefont {Yi}, \citenamefont {Liang}, \citenamefont
  {He}, \citenamefont {He}, \citenamefont {Peng}, \citenamefont {Liu},
  \citenamefont {Liu}, \citenamefont {Zhao}, \citenamefont {Liu}, \citenamefont
  {Dong}, \citenamefont {Zhang}, \citenamefont {Chen}, \citenamefont {Xu},
  \citenamefont {Dai}, \citenamefont {Fang},\ and\ \citenamefont
  {Zhou}}]{feng_strong_2015}%
  \BibitemOpen
  \bibfield  {author} {\bibinfo {author} {\bibfnamefont {Y.}~\bibnamefont
  {Feng}}, \bibinfo {author} {\bibfnamefont {Z.}~\bibnamefont {Wang}}, \bibinfo
  {author} {\bibfnamefont {C.}~\bibnamefont {Chen}}, \bibinfo {author}
  {\bibfnamefont {Y.}~\bibnamefont {Shi}}, \bibinfo {author} {\bibfnamefont
  {Z.}~\bibnamefont {Xie}}, \bibinfo {author} {\bibfnamefont {H.}~\bibnamefont
  {Yi}}, \bibinfo {author} {\bibfnamefont {A.}~\bibnamefont {Liang}}, \bibinfo
  {author} {\bibfnamefont {S.}~\bibnamefont {He}}, \bibinfo {author}
  {\bibfnamefont {J.}~\bibnamefont {He}}, \bibinfo {author} {\bibfnamefont
  {Y.}~\bibnamefont {Peng}}, \bibinfo {author} {\bibfnamefont {X.}~\bibnamefont
  {Liu}}, \bibinfo {author} {\bibfnamefont {Y.}~\bibnamefont {Liu}}, \bibinfo
  {author} {\bibfnamefont {L.}~\bibnamefont {Zhao}}, \bibinfo {author}
  {\bibfnamefont {G.}~\bibnamefont {Liu}}, \bibinfo {author} {\bibfnamefont
  {X.}~\bibnamefont {Dong}}, \bibinfo {author} {\bibfnamefont {J.}~\bibnamefont
  {Zhang}}, \bibinfo {author} {\bibfnamefont {C.}~\bibnamefont {Chen}},
  \bibinfo {author} {\bibfnamefont {Z.}~\bibnamefont {Xu}}, \bibinfo {author}
  {\bibfnamefont {X.}~\bibnamefont {Dai}}, \bibinfo {author} {\bibfnamefont
  {Z.}~\bibnamefont {Fang}},\ and\ \bibinfo {author} {\bibfnamefont {X.~J.}\
  \bibnamefont {Zhou}},\ }\bibfield  {title} {\bibinfo {title} {Strong
  {Anisotropy} of {Dirac} {Cones} in {SrMnBi$_2$} and {CaMnBi$_2$} {Revealed}
  by {Angle}-{Resolved} {Photoemission} {Spectroscopy}},\ }\href
  {https://doi.org/10.1038/srep05385} {\bibfield  {journal} {\bibinfo
  {journal} {Scientific Reports}\ }\textbf {\bibinfo {volume} {4}},\ \bibinfo
  {pages} {5385} (\bibinfo {year} {2015})}\BibitemShut {NoStop}%
\bibitem [{\citenamefont {Borisenko}\ \emph {et~al.}(2019)\citenamefont
  {Borisenko}, \citenamefont {Evtushinsky}, \citenamefont {Gibson},
  \citenamefont {Yaresko}, \citenamefont {Koepernik}, \citenamefont {Kim},
  \citenamefont {Ali}, \citenamefont {van~den Brink}, \citenamefont {Hoesch},
  \citenamefont {Fedorov}, \citenamefont {Haubold}, \citenamefont
  {Kushnirenko}, \citenamefont {Soldatov}, \citenamefont {Schäfer},\ and\
  \citenamefont {Cava}}]{borisenko_time-reversal_2019}%
  \BibitemOpen
  \bibfield  {author} {\bibinfo {author} {\bibfnamefont {S.}~\bibnamefont
  {Borisenko}}, \bibinfo {author} {\bibfnamefont {D.}~\bibnamefont
  {Evtushinsky}}, \bibinfo {author} {\bibfnamefont {Q.}~\bibnamefont {Gibson}},
  \bibinfo {author} {\bibfnamefont {A.}~\bibnamefont {Yaresko}}, \bibinfo
  {author} {\bibfnamefont {K.}~\bibnamefont {Koepernik}}, \bibinfo {author}
  {\bibfnamefont {T.}~\bibnamefont {Kim}}, \bibinfo {author} {\bibfnamefont
  {M.}~\bibnamefont {Ali}}, \bibinfo {author} {\bibfnamefont {J.}~\bibnamefont
  {van~den Brink}}, \bibinfo {author} {\bibfnamefont {M.}~\bibnamefont
  {Hoesch}}, \bibinfo {author} {\bibfnamefont {A.}~\bibnamefont {Fedorov}},
  \bibinfo {author} {\bibfnamefont {E.}~\bibnamefont {Haubold}}, \bibinfo
  {author} {\bibfnamefont {Y.}~\bibnamefont {Kushnirenko}}, \bibinfo {author}
  {\bibfnamefont {I.}~\bibnamefont {Soldatov}}, \bibinfo {author}
  {\bibfnamefont {R.}~\bibnamefont {Schäfer}},\ and\ \bibinfo {author}
  {\bibfnamefont {R.~J.}\ \bibnamefont {Cava}},\ }\bibfield  {title} {\bibinfo
  {title} {Time-reversal symmetry breaking type-{II} {Weyl} state in
  {YbMnBi$_2$}},\ }\href {https://doi.org/10.1038/s41467-019-11393-5}
  {\bibfield  {journal} {\bibinfo  {journal} {Nature Communications}\ }\textbf
  {\bibinfo {volume} {10}},\ \bibinfo {pages} {3424} (\bibinfo {year}
  {2019})}\BibitemShut {NoStop}%
\bibitem [{\citenamefont {Zhao}\ \emph {et~al.}(2018)\citenamefont {Zhao},
  \citenamefont {Golias}, \citenamefont {Zhang}, \citenamefont {Krivenkov},
  \citenamefont {Jesche}, \citenamefont {Gu}, \citenamefont {Rader},
  \citenamefont {Mazin},\ and\ \citenamefont {Gegenwart}}]{zhao_quantum_2018}%
  \BibitemOpen
  \bibfield  {author} {\bibinfo {author} {\bibfnamefont {K.}~\bibnamefont
  {Zhao}}, \bibinfo {author} {\bibfnamefont {E.}~\bibnamefont {Golias}},
  \bibinfo {author} {\bibfnamefont {Q.~H.}\ \bibnamefont {Zhang}}, \bibinfo
  {author} {\bibfnamefont {M.}~\bibnamefont {Krivenkov}}, \bibinfo {author}
  {\bibfnamefont {A.}~\bibnamefont {Jesche}}, \bibinfo {author} {\bibfnamefont
  {L.}~\bibnamefont {Gu}}, \bibinfo {author} {\bibfnamefont {O.}~\bibnamefont
  {Rader}}, \bibinfo {author} {\bibfnamefont {I.~I.}\ \bibnamefont {Mazin}},\
  and\ \bibinfo {author} {\bibfnamefont {P.}~\bibnamefont {Gegenwart}},\
  }\bibfield  {title} {\bibinfo {title} {Quantum oscillations and {Dirac}
  dispersion in the {BaZnBi$_2$} semimetal guaranteed by local {Zn} vacancy
  order},\ }\href {https://doi.org/10.1103/PhysRevB.97.115166} {\bibfield
  {journal} {\bibinfo  {journal} {Physical Review B}\ }\textbf {\bibinfo
  {volume} {97}},\ \bibinfo {pages} {115166} (\bibinfo {year}
  {2018})}\BibitemShut {NoStop}%
\bibitem [{\citenamefont {Thirupathaiah}\ \emph {et~al.}(2019)\citenamefont
  {Thirupathaiah}, \citenamefont {Efremov}, \citenamefont {Kushnirenko},
  \citenamefont {Haubold}, \citenamefont {Kim}, \citenamefont {Pienning},
  \citenamefont {Morozov}, \citenamefont {Aswartham}, \citenamefont
  {Büchner},\ and\ \citenamefont {Borisenko}}]{thirupathaiah_absence_2019}%
  \BibitemOpen
  \bibfield  {author} {\bibinfo {author} {\bibfnamefont {S.}~\bibnamefont
  {Thirupathaiah}}, \bibinfo {author} {\bibfnamefont {D.}~\bibnamefont
  {Efremov}}, \bibinfo {author} {\bibfnamefont {Y.}~\bibnamefont
  {Kushnirenko}}, \bibinfo {author} {\bibfnamefont {E.}~\bibnamefont
  {Haubold}}, \bibinfo {author} {\bibfnamefont {T.~K.}\ \bibnamefont {Kim}},
  \bibinfo {author} {\bibfnamefont {B.~R.}\ \bibnamefont {Pienning}}, \bibinfo
  {author} {\bibfnamefont {I.}~\bibnamefont {Morozov}}, \bibinfo {author}
  {\bibfnamefont {S.}~\bibnamefont {Aswartham}}, \bibinfo {author}
  {\bibfnamefont {B.}~\bibnamefont {Büchner}},\ and\ \bibinfo {author}
  {\bibfnamefont {S.~V.}\ \bibnamefont {Borisenko}},\ }\bibfield  {title}
  {\bibinfo {title} {Absence of {Dirac} fermions in layered {BaZnBi$_2$}},\
  }\href {https://doi.org/10.1103/PhysRevMaterials.3.024202} {\bibfield
  {journal} {\bibinfo  {journal} {Physical Review Materials}\ }\textbf
  {\bibinfo {volume} {3}},\ \bibinfo {pages} {024202} (\bibinfo {year}
  {2019})}\BibitemShut {NoStop}%
\bibitem [{\citenamefont {Schoop}\ \emph {et~al.}(2016)\citenamefont {Schoop},
  \citenamefont {Ali}, \citenamefont {Straßer}, \citenamefont {Topp},
  \citenamefont {Varykhalov}, \citenamefont {Marchenko}, \citenamefont
  {Duppel}, \citenamefont {Parkin}, \citenamefont {Lotsch},\ and\ \citenamefont
  {Ast}}]{schoop_dirac_2016}%
  \BibitemOpen
  \bibfield  {author} {\bibinfo {author} {\bibfnamefont {L.~M.}\ \bibnamefont
  {Schoop}}, \bibinfo {author} {\bibfnamefont {M.~N.}\ \bibnamefont {Ali}},
  \bibinfo {author} {\bibfnamefont {C.}~\bibnamefont {Straßer}}, \bibinfo
  {author} {\bibfnamefont {A.}~\bibnamefont {Topp}}, \bibinfo {author}
  {\bibfnamefont {A.}~\bibnamefont {Varykhalov}}, \bibinfo {author}
  {\bibfnamefont {D.}~\bibnamefont {Marchenko}}, \bibinfo {author}
  {\bibfnamefont {V.}~\bibnamefont {Duppel}}, \bibinfo {author} {\bibfnamefont
  {S.~S.~P.}\ \bibnamefont {Parkin}}, \bibinfo {author} {\bibfnamefont {B.~V.}\
  \bibnamefont {Lotsch}},\ and\ \bibinfo {author} {\bibfnamefont {C.~R.}\
  \bibnamefont {Ast}},\ }\bibfield  {title} {\bibinfo {title} {Dirac cone
  protected by non-symmorphic symmetry and three-dimensional {Dirac} line node
  in {ZrSiS}},\ }\href {https://doi.org/10.1038/ncomms11696} {\bibfield
  {journal} {\bibinfo  {journal} {Nature Communications}\ }\textbf {\bibinfo
  {volume} {7}},\ \bibinfo {pages} {11696} (\bibinfo {year}
  {2016})}\BibitemShut {NoStop}%
\bibitem [{\citenamefont {Liang}\ \emph {et~al.}(2015)\citenamefont {Liang},
  \citenamefont {Gibson}, \citenamefont {Ali}, \citenamefont {Liu},
  \citenamefont {Cava},\ and\ \citenamefont {Ong}}]{liang_ultrahigh_2015}%
  \BibitemOpen
  \bibfield  {author} {\bibinfo {author} {\bibfnamefont {T.}~\bibnamefont
  {Liang}}, \bibinfo {author} {\bibfnamefont {Q.}~\bibnamefont {Gibson}},
  \bibinfo {author} {\bibfnamefont {M.~N.}\ \bibnamefont {Ali}}, \bibinfo
  {author} {\bibfnamefont {M.}~\bibnamefont {Liu}}, \bibinfo {author}
  {\bibfnamefont {R.~J.}\ \bibnamefont {Cava}},\ and\ \bibinfo {author}
  {\bibfnamefont {N.~P.}\ \bibnamefont {Ong}},\ }\bibfield  {title} {\bibinfo
  {title} {Ultrahigh mobility and giant magnetoresistance in the {Dirac}
  semimetal {Cd$_3$As$_2$}},\ }\href {https://doi.org/10.1038/nmat4143}
  {\bibfield  {journal} {\bibinfo  {journal} {Nature Materials}\ }\textbf
  {\bibinfo {volume} {14}},\ \bibinfo {pages} {280} (\bibinfo {year}
  {2015})}\BibitemShut {NoStop}%
\bibitem [{\citenamefont {Liu}\ \emph {et~al.}(2014)\citenamefont {Liu},
  \citenamefont {Jiang}, \citenamefont {Zhou}, \citenamefont {Wang},
  \citenamefont {Zhang}, \citenamefont {Weng}, \citenamefont {Prabhakaran},
  \citenamefont {Mo}, \citenamefont {Peng}, \citenamefont {Dudin},
  \citenamefont {Kim}, \citenamefont {Hoesch}, \citenamefont {Fang},
  \citenamefont {Dai}, \citenamefont {Shen}, \citenamefont {Feng},
  \citenamefont {Hussain},\ and\ \citenamefont {Chen}}]{liu_stable_2014}%
  \BibitemOpen
  \bibfield  {author} {\bibinfo {author} {\bibfnamefont {Z.~K.}\ \bibnamefont
  {Liu}}, \bibinfo {author} {\bibfnamefont {J.}~\bibnamefont {Jiang}}, \bibinfo
  {author} {\bibfnamefont {B.}~\bibnamefont {Zhou}}, \bibinfo {author}
  {\bibfnamefont {Z.~J.}\ \bibnamefont {Wang}}, \bibinfo {author}
  {\bibfnamefont {Y.}~\bibnamefont {Zhang}}, \bibinfo {author} {\bibfnamefont
  {H.~M.}\ \bibnamefont {Weng}}, \bibinfo {author} {\bibfnamefont
  {D.}~\bibnamefont {Prabhakaran}}, \bibinfo {author} {\bibfnamefont {S.-K.}\
  \bibnamefont {Mo}}, \bibinfo {author} {\bibfnamefont {H.}~\bibnamefont
  {Peng}}, \bibinfo {author} {\bibfnamefont {P.}~\bibnamefont {Dudin}},
  \bibinfo {author} {\bibfnamefont {T.}~\bibnamefont {Kim}}, \bibinfo {author}
  {\bibfnamefont {M.}~\bibnamefont {Hoesch}}, \bibinfo {author} {\bibfnamefont
  {Z.}~\bibnamefont {Fang}}, \bibinfo {author} {\bibfnamefont {X.}~\bibnamefont
  {Dai}}, \bibinfo {author} {\bibfnamefont {Z.~X.}\ \bibnamefont {Shen}},
  \bibinfo {author} {\bibfnamefont {D.~L.}\ \bibnamefont {Feng}}, \bibinfo
  {author} {\bibfnamefont {Z.}~\bibnamefont {Hussain}},\ and\ \bibinfo {author}
  {\bibfnamefont {Y.~L.}\ \bibnamefont {Chen}},\ }\bibfield  {title} {\bibinfo
  {title} {A stable three-dimensional topological {Dirac} semimetal
  {Cd$_3$As$_2$}},\ }\href {https://doi.org/10.1038/nmat3990} {\bibfield
  {journal} {\bibinfo  {journal} {Nature Materials}\ }\textbf {\bibinfo
  {volume} {13}},\ \bibinfo {pages} {677} (\bibinfo {year} {2014})}\BibitemShut
  {NoStop}%
\bibitem [{\citenamefont {Neupane}\ \emph {et~al.}(2014)\citenamefont
  {Neupane}, \citenamefont {Xu}, \citenamefont {Sankar}, \citenamefont
  {Alidoust}, \citenamefont {Bian}, \citenamefont {Liu}, \citenamefont
  {Belopolski}, \citenamefont {Chang}, \citenamefont {Jeng}, \citenamefont
  {Lin}, \citenamefont {Bansil}, \citenamefont {Chou},\ and\ \citenamefont
  {Hasan}}]{neupane_observation_2014}%
  \BibitemOpen
  \bibfield  {author} {\bibinfo {author} {\bibfnamefont {M.}~\bibnamefont
  {Neupane}}, \bibinfo {author} {\bibfnamefont {S.-Y.}\ \bibnamefont {Xu}},
  \bibinfo {author} {\bibfnamefont {R.}~\bibnamefont {Sankar}}, \bibinfo
  {author} {\bibfnamefont {N.}~\bibnamefont {Alidoust}}, \bibinfo {author}
  {\bibfnamefont {G.}~\bibnamefont {Bian}}, \bibinfo {author} {\bibfnamefont
  {C.}~\bibnamefont {Liu}}, \bibinfo {author} {\bibfnamefont {I.}~\bibnamefont
  {Belopolski}}, \bibinfo {author} {\bibfnamefont {T.-R.}\ \bibnamefont
  {Chang}}, \bibinfo {author} {\bibfnamefont {H.-T.}\ \bibnamefont {Jeng}},
  \bibinfo {author} {\bibfnamefont {H.}~\bibnamefont {Lin}}, \bibinfo {author}
  {\bibfnamefont {A.}~\bibnamefont {Bansil}}, \bibinfo {author} {\bibfnamefont
  {F.}~\bibnamefont {Chou}},\ and\ \bibinfo {author} {\bibfnamefont {M.~Z.}\
  \bibnamefont {Hasan}},\ }\bibfield  {title} {\bibinfo {title} {Observation of
  a three-dimensional topological {Dirac} semimetal phase in high-mobility
  {Cd$_3$As$_2$}},\ }\href {https://doi.org/10.1038/ncomms4786} {\bibfield
  {journal} {\bibinfo  {journal} {Nature Communications}\ }\textbf {\bibinfo
  {volume} {5}},\ \bibinfo {pages} {3786} (\bibinfo {year} {2014})}\BibitemShut
  {NoStop}%
\bibitem [{\citenamefont {Ortiz}\ \emph {et~al.}(2020)\citenamefont {Ortiz},
  \citenamefont {Teicher}, \citenamefont {Hu}, \citenamefont {Zuo},
  \citenamefont {Sarte}, \citenamefont {Schueller}, \citenamefont {Abeykoon},
  \citenamefont {Krogstad}, \citenamefont {Rosenkranz}, \citenamefont {Osborn},
  \citenamefont {Seshadri}, \citenamefont {Balents}, \citenamefont {He},\ and\
  \citenamefont {Wilson}}]{ortiz_cs_2020}%
  \BibitemOpen
  \bibfield  {author} {\bibinfo {author} {\bibfnamefont {B.~R.}\ \bibnamefont
  {Ortiz}}, \bibinfo {author} {\bibfnamefont {S.~M.}\ \bibnamefont {Teicher}},
  \bibinfo {author} {\bibfnamefont {Y.}~\bibnamefont {Hu}}, \bibinfo {author}
  {\bibfnamefont {J.~L.}\ \bibnamefont {Zuo}}, \bibinfo {author} {\bibfnamefont
  {P.~M.}\ \bibnamefont {Sarte}}, \bibinfo {author} {\bibfnamefont {E.~C.}\
  \bibnamefont {Schueller}}, \bibinfo {author} {\bibfnamefont {A.~M.}\
  \bibnamefont {Abeykoon}}, \bibinfo {author} {\bibfnamefont {M.~J.}\
  \bibnamefont {Krogstad}}, \bibinfo {author} {\bibfnamefont {S.}~\bibnamefont
  {Rosenkranz}}, \bibinfo {author} {\bibfnamefont {R.}~\bibnamefont {Osborn}},
  \bibinfo {author} {\bibfnamefont {R.}~\bibnamefont {Seshadri}}, \bibinfo
  {author} {\bibfnamefont {L.}~\bibnamefont {Balents}}, \bibinfo {author}
  {\bibfnamefont {J.}~\bibnamefont {He}},\ and\ \bibinfo {author}
  {\bibfnamefont {S.~D.}\ \bibnamefont {Wilson}},\ }\bibfield  {title}
  {\bibinfo {title} {{CsV$_3$Sb$_5$}: {A} {$Z_2$} {Topological} {Kagome}
  {Metal} with a {Superconducting} {Ground} {State}},\ }\href
  {https://doi.org/10.1103/PhysRevLett.125.247002} {\bibfield  {journal}
  {\bibinfo  {journal} {Physical Review Letters}\ }\textbf {\bibinfo {volume}
  {125}},\ \bibinfo {pages} {247002} (\bibinfo {year} {2020})}\BibitemShut
  {NoStop}%
\bibitem [{\citenamefont {Nie}\ \emph {et~al.}(2022)\citenamefont {Nie},
  \citenamefont {Sun}, \citenamefont {Ma}, \citenamefont {Song}, \citenamefont
  {Zheng}, \citenamefont {Liang}, \citenamefont {Wu}, \citenamefont {Yu},
  \citenamefont {Li}, \citenamefont {Shan}, \citenamefont {Zhao}, \citenamefont
  {Li}, \citenamefont {Kang}, \citenamefont {Wu}, \citenamefont {Zhou},
  \citenamefont {Liu}, \citenamefont {Xiang}, \citenamefont {Ying},
  \citenamefont {Wang}, \citenamefont {Wu},\ and\ \citenamefont
  {Chen}}]{nie_CDW-driven_2022}%
  \BibitemOpen
  \bibfield  {author} {\bibinfo {author} {\bibfnamefont {L.}~\bibnamefont
  {Nie}}, \bibinfo {author} {\bibfnamefont {K.}~\bibnamefont {Sun}}, \bibinfo
  {author} {\bibfnamefont {W.}~\bibnamefont {Ma}}, \bibinfo {author}
  {\bibfnamefont {D.}~\bibnamefont {Song}}, \bibinfo {author} {\bibfnamefont
  {L.}~\bibnamefont {Zheng}}, \bibinfo {author} {\bibfnamefont
  {Z.}~\bibnamefont {Liang}}, \bibinfo {author} {\bibfnamefont
  {P.}~\bibnamefont {Wu}}, \bibinfo {author} {\bibfnamefont {F.}~\bibnamefont
  {Yu}}, \bibinfo {author} {\bibfnamefont {J.}~\bibnamefont {Li}}, \bibinfo
  {author} {\bibfnamefont {M.}~\bibnamefont {Shan}}, \bibinfo {author}
  {\bibfnamefont {D.}~\bibnamefont {Zhao}}, \bibinfo {author} {\bibfnamefont
  {S.}~\bibnamefont {Li}}, \bibinfo {author} {\bibfnamefont {B.}~\bibnamefont
  {Kang}}, \bibinfo {author} {\bibfnamefont {Z.}~\bibnamefont {Wu}}, \bibinfo
  {author} {\bibfnamefont {Y.}~\bibnamefont {Zhou}}, \bibinfo {author}
  {\bibfnamefont {K.}~\bibnamefont {Liu}}, \bibinfo {author} {\bibfnamefont
  {Z.}~\bibnamefont {Xiang}}, \bibinfo {author} {\bibfnamefont
  {J.}~\bibnamefont {Ying}}, \bibinfo {author} {\bibfnamefont {Z.}~\bibnamefont
  {Wang}}, \bibinfo {author} {\bibfnamefont {T.}~\bibnamefont {Wu}},\ and\
  \bibinfo {author} {\bibfnamefont {X.}~\bibnamefont {Chen}},\ }\bibfield
  {title} {\bibinfo {title} {Charge-density-wave-driven electronic nematicity
  in a kagome superconductor},\ }\href
  {https://doi.org/10.1038/s41586-022-04493-8} {\bibfield  {journal} {\bibinfo
  {journal} {Nature}\ }\textbf {\bibinfo {volume} {604}},\ \bibinfo {pages}
  {59} (\bibinfo {year} {2022})}\BibitemShut {NoStop}%
\bibitem [{\citenamefont {Arachchige}\ \emph {et~al.}(2022)\citenamefont
  {Arachchige}, \citenamefont {Meier}, \citenamefont {Marshall}, \citenamefont
  {Matsuoka}, \citenamefont {Xue}, \citenamefont {McGuire}, \citenamefont
  {Hermann}, \citenamefont {Cao},\ and\ \citenamefont
  {Mandrus}}]{arachchige_charge_2022}%
  \BibitemOpen
  \bibfield  {author} {\bibinfo {author} {\bibfnamefont {H.~W.~S.}\
  \bibnamefont {Arachchige}}, \bibinfo {author} {\bibfnamefont {W.~R.}\
  \bibnamefont {Meier}}, \bibinfo {author} {\bibfnamefont {M.}~\bibnamefont
  {Marshall}}, \bibinfo {author} {\bibfnamefont {T.}~\bibnamefont {Matsuoka}},
  \bibinfo {author} {\bibfnamefont {R.}~\bibnamefont {Xue}}, \bibinfo {author}
  {\bibfnamefont {M.~A.}\ \bibnamefont {McGuire}}, \bibinfo {author}
  {\bibfnamefont {R.~P.}\ \bibnamefont {Hermann}}, \bibinfo {author}
  {\bibfnamefont {H.}~\bibnamefont {Cao}},\ and\ \bibinfo {author}
  {\bibfnamefont {D.}~\bibnamefont {Mandrus}},\ }\bibfield  {title} {\bibinfo
  {title} {Charge {Density} {Wave} in {Kagome} {Lattice} {Intermetallic}
  {ScV$_6$Sn$_6$}},\ }\href {https://doi.org/10.1103/PhysRevLett.129.216402}
  {\bibfield  {journal} {\bibinfo  {journal} {Physical Review Letters}\
  }\textbf {\bibinfo {volume} {129}},\ \bibinfo {pages} {216402} (\bibinfo
  {year} {2022})}\BibitemShut {NoStop}%
\bibitem [{\citenamefont {Tan}\ \emph {et~al.}(2021)\citenamefont {Tan},
  \citenamefont {Liu}, \citenamefont {Wang},\ and\ \citenamefont
  {Yan}}]{tan_charge_2021}%
  \BibitemOpen
  \bibfield  {author} {\bibinfo {author} {\bibfnamefont {H.}~\bibnamefont
  {Tan}}, \bibinfo {author} {\bibfnamefont {Y.}~\bibnamefont {Liu}}, \bibinfo
  {author} {\bibfnamefont {Z.}~\bibnamefont {Wang}},\ and\ \bibinfo {author}
  {\bibfnamefont {B.}~\bibnamefont {Yan}},\ }\bibfield  {title} {\bibinfo
  {title} {Charge {Density} {Waves} and {Electronic} {Properties} of
  {Superconducting} {Kagome} {Metals}},\ }\href
  {https://doi.org/10.1103/PhysRevLett.127.046401} {\bibfield  {journal}
  {\bibinfo  {journal} {Physical Review Letters}\ }\textbf {\bibinfo {volume}
  {127}},\ \bibinfo {pages} {046401} (\bibinfo {year} {2021})}\BibitemShut
  {NoStop}%
\end{thebibliography}
%

\end{document}